\documentclass[11pt,a4paper]{article}

\usepackage{jcappub} 

\usepackage[T1]{fontenc} 

\title{\boldmath On gravitational baryogenesis in inflationary$f(R)$ cosmologies}
 
\author{Ian B. Whittingham}

\affiliation{College of Science and Engineering,
James Cook University, \\ Townsville, Queensland, Australia 4811}

\emailAdd{Ian.Whittingham@jcu.edu.au}

\abstract{The origin of the small asymmetry in the abundance of matter and antimatter in the very early universe, described by the 
Baryon Asymmetry Factor $\eta =8.65 \times 10^{-11}$, is one of the big unanswered questions of contemporary physics. 

Gravitational baryogenesis postulates the baryon-antibaryon asymmetry is generated by a coupling between the baryon current 
and the gradient of the Ricci scalar $R$. As this coupling is zero in General Relativity for a radiation dominated flat early 
Universe, more generalized theories of gravity are required. 

The present study investigates the feasibility of gravitational baryogenesis during the inflationary  phase of two $f(R)$ 
modified theories of gravity: the widely used Starobinsky $f(R)=R+R^{2}/M^{2}$ model, and the recently proposed power-law model 
$f(R)=c_{1}R^{2+k/4}+c_{2}R+c_{3}$ of Odintsov and Oikonomou (2025) that is constructed from the slow-roll inflation parameters, 
and fits the new high-multipole CMB observations reported by the Planck and ACT collaborations for $k \sim -0.03$. 

Rather than use the Jordan frame in which these cosmologies have been formulated, the present investigation is undertaken 
in the Einstein frame, which is obtained from the Jordan frame by a conformal transformation $\Omega$, which 
introduces the scalaron field $\phi=\kappa^{-1} \sqrt{6} \ln \Omega$. 
The motion of the scalaron is studied for the slow roll inflationary era using analytic approximate solutions, from which expressions are obtained for $R$ and hence $\eta$. Some of the expressions for the Starobinsky model were obtained first by Motohashi and Nishizawa (2012) but the expressions for the power-law model of Odintsov and Oikonomou and its generalization are new. 

Although the values of $\eta$ calculated for both models are quite close to the observed value, they are based upon several assumptions that qualify these results. The asymmetries are generated during the slow roll inflation era and, for values generated $N$ e-folds before the end of inflation, must be diluted by the factor $\Delta \lesssim 10\, e^{-3N}$, making agreement with observation difficult to achieve unless baryogenesis occurs at the end of the inflation era. The calculations also assume rapid thermalization of the plasma of the gravitationally produced particles, and that the interaction rate of the baryon-violating processes is greater than the expansion rate of the Universe until decoupling. The present calculations strongly violate these assumptions and would require significant variation of input parameters, which is excluded by CMB observations, to obtain agreement. 

A complete treatment of thermalization of the gravitationally produced plasma, baryon-number violating freeze-out, 
and entropy dilution is required to establish whether or not gravitational baryogenesis during slow-roll inflation is a viable 
mechanism for producing the observed baryon asymmetry of the Universe for these inflationary $f(R)$ cosmologies.} 


\arxivnumber{2602.19075}

\begin{document}
\maketitle
\flushbottom


\section{Introduction}

Big Bang Nucleosynthesis (BBN), and measurements of the anisotropies of the Cosmic Microwave Background (CMB) 
combined with the large scale structure of the Universe, show that the number density of matter is of the order of $10^{-9}$
of that of the photon number density. As there is no antimatter in the present Universe other than that produced in cosmic rays,
this indicates that matter is dominant over antimatter.
This asymmetry is represented by the Baryon Asymmetry Factor (BAF)
\begin{equation}
\label{IW0}
\eta \equiv \frac{n_{\mathrm{B}}-n_{\bar{\mathrm{B}}}}{s},
\end{equation}
where $n_{\mathrm{B}}(n_{\bar{\mathrm{B}}})$ is the number density of baryons (antibaryons), and $s$ is the entropy density of the relativistic radiation bath. For the present Universe $s$ is related to the photon number density 
$n_{\gamma}$ via $s=7.040\, n_{\gamma}$. The entropy density is a better quantity to compare the baryon number densities
to than the photon density as heavy particles that were in thermal equilibrium in the early Universe annihilated as the Universe
cooled and produced more photons than baryons~\cite{Cline2006}. 
Current observations~\cite{Cooke2014,Ade2016} give
\begin{eqnarray}
\label{IW1}
\eta \equiv \frac{n_{\mathrm{B}}-n_{\bar{\mathrm{B}}}}{s} & =&\frac{1}{7.04} \frac{n_{\mathrm{B}}-n_{\bar{\mathrm{B}}}}{n_{\gamma}}
\nonumber  \\
& = &\left\{
			\begin{array}{ll}
			(8.2-9.4) \times 10^{-11} & (\mathrm{BBN}) \\
			(8.65 \pm 0.09) \times 10^{-11} & (\mathrm{CMB})
			\end{array} \right.
\end{eqnarray}

Conventionally, this asymmetry is generated dynamically from an initial baryon-symmetric phase in the primordial Universe. 
Sakharov~\cite{Sak1967} established the necessary conditions for such a generation of baryon asymmetry:
\begin{enumerate}
\item
Baryon-number non-conserving interactions.
\item
$C$ and $CP$ violation so that processes generating baryons are more rapid than those generating antibaryons.
\item
Departure from thermal equilibrium since, from $CPT$ symmetry, the equilibrium phase space densities of particles and 
antiparticles are the same. Any reaction producing $n_{\mathrm{B}} \neq n_{\bar{\mathrm{B}}}$ must freeze out before 
particles and antiparticles achieve thermodynamical equilibrium.
\end{enumerate}
The first two conditions relate to the particle physics interactions present but the departure from thermal equilibrium is gravitational and connects the microscopic and macroscopic scales~\cite{PFLM2023}. 

One popular mechanism for departure from thermal equilibrium is the electroweak phase transition which occurs during the expansion of the Universe at $T \sim 100$ GeV. This has been studied both within the Standard Model~\cite{Vis2025} and, particularly, its extension to include an additional Higgs doublet, 
possibly arising from supersymmetry. Baryon and lepton violation is generated by transitions between different topological
sectors of the gauge and Higgs fields, which are separated by energy barriers corresponding to saddle point sphaleron configurations
(see, eg, \cite{ACW1993}).

Gravitational Baryogenesis~\cite{DKKMS2004} provides a mechanism for generating a baryon asymmetry during the expansion of the universe by means of a dynamical breaking of $CPT$ (and $CP$) in which thermal equilibrium is preserved. It postulates a gravitational interaction between the baryon number current $j^{\mu}_{\mathrm{B}}$ and the derivative $\partial_{\mu} R$ of the Ricci scalar $R$, with the action
\begin{eqnarray}
\label{IW2}
\mathcal{S}_{\mathrm{GB}} & = & \frac{1}{M_{\ast}^{2}} \int d^{4}x \sqrt{-g} (\partial _{\mu}R) j^{\mu}_{\mathrm{B}} \\
                          & \equiv & \int d^{4}x \sqrt{-g} \mathcal{L}_{\mathrm{GB}}.
\end{eqnarray}
This interaction biases the dynamics in the expanding Universe, causing an asymmetry between 
particles and antiparticles. It requires the $B$-violating processes to be in thermal equilibrium. 
The interaction modifies the thermal equilibrium distributions in a similar fashion to a chemical potential. 
$B$-violation decouples below a temperature $T_{\mathrm{D}}$. The baryon asymmetry must be generated after the main
release of entropy during inflation~\cite{EW2023}.

Davoudiasl et al.~\cite{DKKMS2004} argue that an interaction of the form $\mathcal{L}_{\mathrm{GB}}$ is expected in the low-energy effective field theory of quantum gravity if the cutoff scale $M_{\ast}$ of the effective theory is of order of the reduced Planck scale:
\begin{equation}
\label{IW3}
M_{\mathrm{Pl}} = (\hbar c/(8 \pi G_{\mathrm{N}}))^{1/2} = 2.435 \times 10^{18} \, \mathrm{GeV}/c^{2}.
\end{equation}
$\mathcal{L}_{\mathrm{GB}}$ can also occur in supergravity theories from a higher-dimensional operator in the K\aa hler potential.

For an expanding spatially uniform homogeneous universe $\partial_{0}R \neq 0$ and $\partial_{i}R =0$, and the interaction reduces to
\begin{equation}
\label{IW4}
\mathcal{L}_{\mathrm{GB}} = -\frac{1}{M_{\ast}^{2}} \partial_{0}R \, j^{0}_{\mathrm{B}}.
\end{equation}

The baryonic current is 
\begin{equation}
\label{IW6}
j^{\mu}_{\mathrm{B}}(x) = q_{\mathrm{B}} \bar{\psi}(x) \gamma^{\mu} \psi(x),
\end{equation}
where $q_{\mathrm{B}}$ is the charge associated with the current. Noting that
\begin{equation}
\label{IW7}
 j^{0}_{\mathrm{B}} = q_{\mathrm{B}} \bar{\psi}(x) \gamma^{0} \psi(x) = q_{\mathrm{B}}\psi^{\dag}(x) \psi(x),
\end{equation}
where $\psi^{\dag}(x) \psi(x)$ is the number density of the baryons, and that $q_{\mathrm{B}} = +1 (-1)$ for baryons (antibaryons),
then
\begin{equation}
\label{IW8}
j^{0}_{\mathrm{B}} = n_{\mathrm{B}} - n_{\bar{\mathrm{B}}}.
\end{equation}

For a system of ultrarelativistic fermions in thermal equilibrium at temperature $T$, their particle number density, given by 
the Fermi-Dirac distribution, is
\begin{equation}
\label{IW35}
n_{\mathrm{B}} - n_{\bar{\mathrm{B}}}
= \frac{g_{\mathrm{B}}}{6 \pi^{2}} \left( \frac{k_{\mathrm{B}}T}{\hbar c} \right)^{3} \,
		\left[ \frac{\mu}{k_{\mathrm{B}}T}\,\pi^{2} + \left(\frac{\mu}{k_{\mathrm{B}}T}\right)^{3}\right],
\end{equation}
where $\mu$ is their chemical potential and $g_{\mathrm{B}}=2$ is their number of spin states.

The interaction (\ref{IW4}) has the form
\begin{equation}
\label{IW37}
\mathcal{L}_{\mathrm{GB}} = \mu \,\delta n_{\mathrm{b}},
\end{equation}
where
\begin{equation}
\label{IW38}
\delta n_{\mathrm{b}} \equiv n_{\mathrm{B}} - n_{\bar{\mathrm{B}}}
\end{equation}
and
\begin{equation}
\label{IW39}
\mu \equiv - \frac{1}{M_{\ast}^{2}} \partial_{0}R
\end{equation}
is the effective chemical potential for the system.

The entropy density of the early Universe for all particles in thermal equilibrium at temperature $T$ is
\begin{equation}
\label{IW40}
s = \frac{k_{\mathrm{B}}^{4}}{(\hbar c)^{3}}\, \frac{2 \pi^{2}}{45} g_{\ast}(T) T^{3},
\end{equation}
where $g_{\ast}(T)$ is the number of degrees of freedom of ultrarelativistic particles present, and is 
equal to 106.75 for $k_{\mathrm{B}}T \gg m_{t}c^{2}$.

The Baryon Asymmetry Factor is therefore
\begin{equation}
\label{IW41}
\eta = \frac{\delta n_{\mathrm{b}}}{s} = \frac{g_{\mathrm{B}}}{6 k_{\mathrm{B}}g_{\ast}(T)}
\frac{45}{2 \pi^{2}} \,\frac{\mu}{k_{\mathrm{B}}T}
\left[1+ \frac{1}{\pi^{2}}\left(\frac{\mu}{k_{\mathrm{B}}T}\right)^{2}\right]
\end{equation}
which, to order $\mu$, is
\begin{equation}
\label{IW42}
\eta = -\left( \frac{15 g_{\mathrm{B}}}{4 \pi^{2} k_{\mathrm{B}}g_{\ast}(T)} \right)
\;\frac{\partial_{0}R}{M_{\ast}^{2}T}.
\end{equation}
Once $T$ drops below the decoupling temperature $T_{\mathrm{D}}$, this ratio is frozen in at the value $\eta|_{T=T_{\mathrm{D}}}$.

In General Relativity $R$ and $\partial_{0}R$ are zero for a flat radiation-dominated early Universe, so more generalized 
theories of gravity such as $f(R)$ gravity need to be considered in order for gravitational baryogenesis to be possible.

The structure of this paper is as follows. A summary of the basic equations of $f(R)$ gravity formulated in the Jordan frame,
its equivalence via a conformal transformation $\Omega$ to a scalar-tensor theory in the Einstein frame, and the basic equations of the scalaron field $\phi = \kappa^{-1} \sqrt{6}\ln \Omega$ in the Einstein frame, are presented in Sec. II. Inflation and reheating 
in $f(R)$ cosmologies and the cosmological gravitational production of particles is discussed in Sec. III, 
and a brief review of existing calculations of gravitational baryogenesis in Sec. IV. 
Implications of recent measurements of the CMB are considered in Sec. V. In Sec. VI, scalaron calculations are presented of 
gravitational baryogenesis during the inflationary era of two $f(R)$ theories of gravity, the widely used Starobinsky 
$f(R)=R+R^{2}/M^{2}$ model, and the recently proposed power-law model $f(R)=c_{1}R^{2+k/4}+c_{2}R+c_{3}$ of Odintsov and Oikonomou. 
The calculations are based on analytic approximate solutions for the scalaron during the inflationary slow-roll era. These expressions for the Starobinsky model were obtained first by Motohashi and Nishizawa (2012), but the expressions for the power-law model of Odintsov and Oikonomou are new. Sect. VII contains a summary and conclusions for the investigation.

\section{$f(R)$ cosmologies}

\subsection{Basic equations}

The $f(R)$ class of modified theories of gravity~\cite{SF2010,NO2011,NOO2017} attempt to explain the current accelerated 
expansion of the Universe, the Dark Matter behaviour, and primordial inflation, that is, 
they seek explanations in terms of gravitational dynamics. The aim is to produce two periods of acceleration (inflation and late-time),
between which there should be a period of decelerated expansion allowing the conventional eras of radiation dominance
and matter dominance.

Motivation also comes from attempts to quantize General Relativity. General Relativity is not renormalizable and therefore
cannot be conventionally quantized. In 1962 \cite{UdeWitt1962} showed that renormalization at one-loop demands the Einstein-Hilbert action be supplemented by higher-order curvature terms.
Also, quantum corrections and string theory require the effective low-energy gravitational action to include 
higher-order curvature invariants~\cite{Stelle1977,Vilk1992,BOS1992}.

Einstein's equations of General Relativity 
\begin{equation}
\label{IW43}
R_{\mu \nu} -\frac{1}{2} g_{\mu \nu} R = \kappa^{2} T_{\mu \nu}
\end{equation}
can be obtained from the Einstein-Hilbert action
\begin{equation}
\label{IW46}
\mathcal S _{\mathrm{EH}} =  \int d^{4}x \sqrt{-g} [\frac{1}{2 \kappa^{2}} R +\mathcal L_{m}(g_{\mu \nu},\psi_{m})]
\end{equation}
by variation with respect to  the metric $g_{\mu \nu}$. Here $\kappa^{2}=8\pi G_{\mathrm{N}} = 1/M_{\mathrm{Pl}}^{2}$, 
$g=\det (g_{\mu \nu})$, $\mathcal L_{m}$ is Lagrangian density for the matter fields
$\psi_{m}(x)$, $R_{\mu \nu}$ is the Ricci curvature tensor
\begin{equation}
\label{IW44}
R_{\mu \nu} = \partial_{\lambda}\Gamma^{\lambda}_{\mu \nu} -\partial_{\nu}\Gamma^{\lambda}_{\mu \lambda}
+\Gamma^{\rho}_{\mu \nu}\Gamma^{\lambda}_{\lambda \rho} - \Gamma^{\rho}_{\mu \lambda} \Gamma^{\lambda}_{\nu \rho},
\end{equation}
$R = R^{\mu}_{\mu}$ is the Ricci scalar, $\Gamma^{\lambda}_{\mu \nu}$ are the affine connections of the metric,
and $T_{\mu \nu}$ is the matter energy-momentum tensor.
We use the Lorentzian signature $(-1,+1,+1,+1)$ and the units $\hbar =c=k_{\mathrm{B}}=1$.  

The $f(R)$ modified theories of gravity assume the generalized action
\begin{equation}
\label{IW48}
\mathcal S  = \int d^{4}x \sqrt{-g} [\frac{1}{2 \kappa^{2}}f(R) +\mathcal L_{m}(g_{\mu \nu},\psi_{m})].
\end{equation}
Metric $f(R)$ gravity is obtained by variation of this action with respect to the metric, to give the field equations
\begin{equation}
\label{IW59}
-\frac{1}{2} g_{\mu \nu}f(R)  +  f^{\prime}(R) R_{\mu \nu}+\nabla_{\rho}\nabla^{\rho} f^{\prime}(R) g_{\mu \nu} 
- \nabla_{\mu} \nabla _{\nu} f^{\prime}(R) - \kappa ^{2} T_{\mu \nu} =0,
\end{equation}
where $f^{\prime}(R)= df/dR$ and 
\begin{equation}
\label{IW47}
T_{\mu \nu } = \frac{-2}{\sqrt{-g}} \frac{\delta (\sqrt{-g}\mathcal L_{m})}{\delta g^{\mu \nu}} .
\end{equation}
Defining
\begin{equation}
\label{IW60}
G_{\mu \nu} \equiv  -\frac{1}{2} g_{\mu \nu}f(R) + f^{\prime}(R) R_{\mu \nu}
+\nabla_{\rho}\nabla^{\rho} f^{\prime}(R) g_{\mu \nu} -\nabla_{\mu} \nabla _{\nu} f^{\prime}(R), 
\end{equation}
then the equations of $f(R)$ gravity are
\begin{equation}
\label{IW61}
G_{\mu \nu } = \kappa^{2} \,T_{\mu \nu}.
\end{equation}

The trace of (\ref{IW61}) gives
\begin{equation}
\label{IW61b}
f^{\prime}(R) R -2 f(R) + 3 \nabla _{\mu} \nabla ^{\mu} f^{\prime}(R) = \kappa^{2} T,
\end{equation}
where $T$ is the trace of the energy-momentum tensor.
$R$ and $T$  are therefore related differentially, in contrast to General Relativity where $R=-\kappa^{2}T$ and
they are related algebraically.

For a flat spatially homogeneous and isotropic universe described by the FRW metric 
\begin{equation}
\label{IW65}
ds^{2} = g_{\mu \nu } dx^{\mu} dx^{\nu} = -dt^{2} +a^{2}(t) [dx^{2}+dy^{2}+dz^{2}],
\end{equation}
the Ricci scalar is 
\begin{equation}
\label{IW69}
R^{\mu}_{\mu} = g^{\mu \nu} R_{\mu \nu} = \frac{6}{a^{2}} [a \ddot{a} + \dot{a}^{2}]
\end{equation}
and, assuming a perfect fluid,
\begin{equation}
\label{IW70}
T_{\mu \nu} = (\rho + p)u_{\mu} u_{\nu} + p g_{\mu \nu}
\end{equation}
where
\begin{equation}
\label{IW74}
T_{00} = \rho, \quad T_{ij}=p \delta_{i,j}.
\end{equation}
and
\begin{equation}
\label{IW75}
T = T^{\mu}_{\mu} = g^{\mu \nu} T_{\mu \nu} = (-\rho + 3p).
\end{equation}

Since
\begin{equation}
\label{IW91}
\nabla^{\mu} G_{\mu \nu} = 0
\end{equation}
then the field equations (\ref{IW61}) give
\begin{equation}
\label{IW93}
\nabla^{\mu} T_{\mu \nu} = 0
\end{equation}
and hence the conservation law for a perfect fluid
\begin{equation}
\label{IW97}
\dot{\rho} +  3 \frac{\dot{a}}{a} (\rho + p) =0.
\end{equation}

In terms of the Hubble parameter
\begin{equation}
\label{IW101}
H \equiv \frac{\dot{a}}{a}
\end{equation}
the $(0,0)$ components of the field equations are
\begin{equation}
\label{IW103}
3 H^{2} f^{\prime}(R) +  \frac{1}{2} [f(R)-R f^{\prime}(R)] + 3 H \dot{R} f^{\prime \prime}(R)
=  \kappa^{2} \rho,
\end{equation}
and the $(i,j)$ components are
\begin{equation}
\label{IW106}
-[2 \dot{H}+ 3 H^{2}]f^{\prime}(R)  -  \frac{1}{2}[f(R)-R f^{\prime}(R)] -\dot{R}^{2} f^{\prime\prime\prime}(R)
- \ddot{R}f^{\prime\prime}(R) -2 H \dot{R} f^{\prime\prime}(R)
= \kappa^{2} p.
\end{equation}

The Trace equation (\ref{IW61b} becomes
\begin{equation}
\label{IW83}
f^{\prime}(R) R - 2 f(R) +3 [- f^{\prime \prime \prime}(R) \dot{R}^{2} - f^{\prime \prime}(R) \ddot{R}
-3 H f^{\prime \prime}(R) \dot{R}]
= \kappa^{2}(3p-\rho).
\end{equation}

\subsection{Equivalence with scalar-tensor Models}

The metric tensor contains spin-2 modes, and vector and scalar modes. General Relativity contains only the 
massless graviton but, when the action includes terms that depend on $R$, $R_{\mu\nu}R^{\mu\nu}$
and $R_{\mu\nu\rho\sigma}R^{\mu\nu\rho\sigma}$, other modes appear. $f(R)$ gravity includes a massive 
scalar mode as it is equivalent to a scalar-tensor theory~\cite{Sot2006}.

To see this, the Jordan and Einstein frames are introduced, which are related by the conformal 
transformation
\begin{equation}
\label{IW182}
g_{\mu\nu}^{\mathrm{J}} = \Omega^{-2} g_{\mu\nu}^{\mathrm{E}},
\end{equation}
where the Jordan frame is the frame in which the $f(R)$ gravity has been formulated, and
\begin{equation}
\label{IW185}
\Omega^{2} = \frac{df}{dR}.
\end{equation}
Defining the scalar field $\phi$ via
\begin{equation}
\label{IW189}
\ln \Omega \equiv \frac{\kappa}{\sqrt{6}}\phi ,
\end{equation}
then the action in the Einstein frame is
\begin{eqnarray}
\label{IW193}
\mathcal S & = & \int d^{4}x \sqrt{-g_{\mathrm{E}}}\left[\frac{1}{2 \kappa^{2}} R_{\mathrm{E}} 
- \frac{1}{2} g^{\mu\nu}_{\mathrm{E}} \partial_{\mu}\phi \,\partial_{\nu}\phi -U(\phi)\right]
\nonumber  \\
& & +\int d^{4}x [e^{-2\sqrt{2/3}\kappa \phi} \sqrt{-g_{\mathrm{E}}}\,] 
\mathcal L_{m}(e^{-\sqrt{2/3}\kappa \phi}g^{\mathrm{E}}_{\mu \nu},\psi_{m}),
\end{eqnarray} 
where the scalar potential is
\begin{equation}
\label{IW192}
U(\phi) \equiv \frac{R(\Omega) \Omega^{2} - f[R(\Omega)]}{2 \kappa^{2} \Omega^{4}}.
\end{equation}
The scalar field $\phi$, called the \textit{scalaron}, couples minimally to the Ricci curvature and has a 
canonical kinetic energy. The Einstein frame perspective of $f(R)$ gravity is therefore the standard General Relativity 
description of gravity supplemented with a canonical scalar field that couples only to matter.

The equations of $f(R)$ gravity in the Einstein frame are obtained by noting that, for a flat FRW metric,
\begin{equation}
\label{IW122c}
dt_{\mathrm{J}}= \Omega^{-1} dt_{\mathrm{E}}, \quad a_{\mathrm{J}} = \Omega^{-1}a_{\mathrm{E}},
\end{equation}
the relationship between the Hubble parameters is
\begin{equation}
\label{IW122h}
H_{\mathrm{J}} = \Omega \left[H_{\mathrm{E}} - \frac{\kappa}{\sqrt{6}}\frac{d\phi}{dt_{\mathrm{E}}}\right],
\end{equation}
and that the matter energy-momentum tensors are related by
\begin{equation}
\label{IW122i}
T_{\mu \nu}^{\mathrm{J}}  =  \Omega^{2} T_{\mu \nu }^{\mathrm{E}}.
\end{equation}

The modified Friedmann equation (\ref{IW103}) then becomes
\begin{eqnarray}
\label{IW103e}
3 H_{\mathrm{E}}^{2} & = &\kappa^{2}\left[\frac{1}{2} \left(\frac{d\phi}{dt_{\mathrm{E}}}\right)^{2} +
U(\phi) + \rho_{\mathrm{E}} \right]
\nonumber  \\
& \equiv & \kappa^{2} [\rho_{\mathrm{E}}^{\phi} + \rho_{\mathrm{E}}],
\end{eqnarray}
and the trace equation (\ref{IW83}) provides the equation of motion for the scalar field 
\begin{equation}
\label{IW83e}
\frac{d^{2}\phi}{dt_{\mathrm{E}}^{2}} + 3H_{\mathrm{E}} \frac{d\phi}{dt_{\mathrm{E}}} + \frac{dU(\phi)}{d\phi}
= \frac{\kappa}{6}(\rho_{\mathrm{E}} - 3 p_{\mathrm{E}}).
\end{equation}

Finally, the energy conservation condition (\ref{IW97})
\begin{equation}
\label{IW83f}
\frac{d\rho^{\mathrm{tot}}_{\mathrm{E}}}{dt_{\mathrm{E}}} = 
-3 H_{\mathrm{E}}(\rho^{\mathrm{tot}}_{\mathrm{E}} +p^{\mathrm{tot}}_{\mathrm{E}}),
\end{equation}
where
\begin{equation}
\label{IW83g}
\rho^{\mathrm{tot}}_{\mathrm{E}} = \rho^{\mathrm{\phi}}_{\mathrm{E}} + \rho_{\mathrm{E}},
\quad p^{\mathrm{tot}}_{\mathrm{E}} = p^{\mathrm{\phi}}_{\mathrm{E}} + p_{\mathrm{E}}
\end{equation}
and
\begin{equation}
\label{IW83h}
\rho^{\mathrm{\phi}}_{\mathrm{E}} = \frac{1}{2} \left(\frac{d \phi}{dt_{\mathrm{E}}}\right)^{2} + U(\phi),
\quad p^{\mathrm{\phi}}_{\mathrm{E}} = \frac{1}{2} \left(\frac{d \phi}{dt_{\mathrm{E}}}\right)^{2} - U(\phi),
\end{equation}
give
\begin{equation}
\label{IW83j}
\frac{dH_{\mathrm{E}}}{dt_{\mathrm{E}}} = -\frac{\kappa^{2}}{2} \left[\left(\frac{d \phi}{dt_{\mathrm{E}}}\right)^{2}
+\rho_{\mathrm{E}}+ p_{\mathrm{E}}\right].
\end{equation}

The field equations (\ref{IW103}) and (\ref{IW83}) in the Jordan frame are 4th-order differential equations in the 
scale factor $a(t_{\mathrm{J}})$, which makes their physical interpretation unclear~\cite{MN2012}. Studying the evolution 
of the scalaron in the Einstein frame clarifies the physical picture and understanding the dynamics.

Hereafter, until Sec. VI, the subscript $\mathrm{J}$ for the Jordan frame will be omitted.

\subsection{Starobinsky $R^{2}$ model}

The Starobinsky model~\cite{Star1980} is defined by
\begin{equation}
\label{IW194}
f(R) = R + \frac{R^{2}}{6\tilde{M}^{2}}.
\end{equation}
Since
\begin{equation}
\label{IW196}
\Omega^{2} = \frac{df(R)}{d R} = 1 + \frac{R}{3\tilde{M}^{2}},
\end{equation}
the scalar potential (\ref{IW192}) is 
\begin{equation}
\label{IW199}
U(\phi) = \frac{3 \tilde{M}^{2}}{4 \kappa^{2}} (1-\Omega^{-2})^{2}
= \frac{3 \tilde{M}^{2}}{4 \kappa^{2}} \left[e^{-\sqrt{2/3}\kappa \phi}-1\right]^{2}.
\end{equation}

No-scale models of inflation~\cite{Ellis2020} provide field-theoretical models that reproduce, at least approximately, the 
Starobinsky model. No-scale supergravity with SO(10) and flipped SU(5)xU(1) GUT models have been studied extensively and, 
recently~\cite{Ellis2025}, SU(5) and SO(10) with pure gravity mediation for soft SUSY breaking, have been advocated.

The $R^{2}$ model is essentially a large-field inflation, with its exponential coupling $\sim \exp(-\alpha \kappa \phi)$ 
naturally arising from the conformal transformation of the metric from the Jordan frame into the Einstein frame. The constant
$\alpha=\sqrt{2/3}$ is uniquely determined by the canonical normalization of the scalaron $\phi$ in the Einstein frame.

\subsection{$R^{2}$-corrected models}

The Starobinsky model is plagued with theoretical problems~\cite{ABS2010} and there have been several attempts to 
overcome these issues.

\cite{OOS2023} study a power law $f(R)$ gravity with an early dark energy term, that can describe both the early-time and late-time acceleration of the Universe. They consider the $f(R)$ gravity function
\begin{equation}
\label{IW210}
f(R) = R  + \frac{R^{2}}{M^{2}} -2 \Lambda \gamma \left(\frac{R}{2 \Lambda}\right)^{\delta} + f_{\mathrm{EDE}},
\end{equation}
where $f_{\mathrm{inf}} \equiv R^{2}/M^{2}$ generates the early-time acceleration and dominates during the 
inflationary era. The constant $M$ should be large enough that $f_{\mathrm{inf}}$ becomes negligible at late times $z < 3000$, 
corresponding to our observational data. It can be evaluated as $M=1.5 \times 10^{-5} (50/N)M_{\mathrm{Pl}}$,
where $N$ is the number of $e$-folds during inflation, that is $M$ is of order $10^{13}$ GeV.

The early dark energy term $f_{\mathrm{EDE}}$ must  lead to regular evolution without singularities. A viable form
is~\cite{OGS2021}
\begin{equation}
\label{IW211}
f_{\mathrm{EDE}} = - 2 \Lambda \alpha \frac{R}{R_{0}+R},
\end{equation}
where $R_{0}$ is the Ricci scalar value for the intermediate epoch $1000 < z < 3000$. A sufficiently
large $\alpha$ can effectively suppress oscillations arising in this model during the intermediate epoch
and beyond. At early times, that is the inflationary era, where $R \gg R_{0}$, this term becomes the constant 
$f_{\mathrm{EDE}} \approx - 2 \alpha\Lambda$
and is irrelevant in the very early Universe. At late times $R \ll R_{0}$, this term becomes small due to the factor $R/R_{0}$.

\cite{ABS2010} have constructed their $gR^{2}-AB$ model by extending the condition $f^{\prime}(R) > 0$ 
for $R > 0$ to $R < 0$.  Their $f(R)$ function is
\begin{equation}
\label{IW209a}
f(R) =  (1-g)R + \frac{R^{2}}{6 \tilde{M}^{2}}
+ g \tilde{M}^{2} \delta \ln \left[\frac{\cosh(R/\delta \tilde{M}^{2} -b)}{\cosh b}\right],
\end{equation}
which can be rewritten as~\cite{MN2012}
\begin{equation}
\label{IW209b}
f(R) =  R  - \frac{R_{\mathrm{vac}}}{2} + \frac{R^{2}}{6 \tilde{M}^{2}}
+ g \tilde{M}^{2} \delta \ln \left[1+e^{-2(R/\delta \tilde{M}^{2} -b)}\right],
\end{equation}
where
\begin{equation}
\label{IW209c}
R_{\mathrm{vac}} \equiv 2 g \tilde{M}^{2} \delta [b+\ln(2 \cosh b)]
\end{equation}
plays the role of the cosmological constant.

The model satisfactorily describes inflation (primordial DE), the present acceleration of the Universe (present DE), and the 
intermediate epochs of radiation and matter domination, for a unique choice of its parameter $\tilde{M}$, determined by the 
observed power of scalar (density) perturbations.

\section{Inflation and reheating in $f(R)$ cosmologies}

Inflation is most commonly modelled as a scalar field cosmology, which postulates a canonical scalar field 
$\phi$ (called the \textit{inflaton}), with a self-interaction potential $V(\phi)$ which has the 
density and pressure (\ref{IW83h}) of a perfect fluid.

However, the introduction of a scalar field to produce a de Sitter phase of inflationary expansion, and to then transfer 
its energy to Standard Model particles by direct couplings to these fields, lacks any fundamental justification. Also,
to obtain a sufficiently large inflation, suitable reheating after inflation, and to make the material fluctuations
small enough to be consistent with observations, requires fine tuning of the relevant couplings or masses~\cite{MMS1986}.

Starobinsky~\cite{Star1980} has suggested an alternate model in which the de Sitter phase is obtained as a self-consistent solution of
the vacuum Einstein's equations modified by the one-loop contributions of conformally covariant quantized matter fields. Thus the 
expansion rate depends on the number of elementary fields in the model. No classical matter fields are involved.
Just like the inflaton model, the de Sitter phase is unstable and has a finite lifetime. However, unlike the inflaton model where 
particle creation arises from coherent oscillations of the inflaton field about the quadratic minimum of the scalar field potential, 
particle creation in the Starobinsky model is due to the rapidly changing spacetime metric as the Universe makes a transition from the 
de Sitter era to either a radiation-dominated or matter-dominated era. 

Einstein's vacuum equations with inclusion of one-loop quantum corrections are~\cite{Vilenkin1985}
\begin{equation}
\label{IWg1}
R_{\mu\nu} - \frac{1}{2} g_{\mu\nu} R = - 8 \pi G_{\mathrm{N}} \langle T_{\mu\nu} \rangle,
\end{equation}
where $T_{\mu\nu}$ is the stress-energy tensor for the theory's various quantum fields and a FRW metric
has been assumed. The quantum corrections are particularly simple for the case of free, massless, conformally invariant 
fields~\cite{BD1982}:
\begin{equation}
\label{IWg3}
\langle T_{\mu\nu} \rangle = k_{1} {}^{(1)}H_{\mu\nu} + k_{3} {}^{(3)}H_{\mu\nu},
\end{equation}
where the tensor ${}^{(1)}H_{\mu\nu}$ is identically conserved and can be obtained by varying the local action
\begin{equation}
\label{IWg6}
{}^{(1)}H_{\mu\nu} = \frac{2}{\sqrt{-g}} \frac{\delta}{\delta g^{\mu\nu}} \int d^{4}x \sqrt{-g} R^{2}.
\end{equation}
The coefficient $k_{1}$ can take any value and must be determined by experiment.

The tensor ${}^{(3)}H_{\mu\nu}$ is conserved only in conformally flat spacetimes (in particular in FRW spacetimes)
and cannot be obtained by varying a local action. Its coefficient $k_{3}$ is uniquely determined:
\begin{equation}
\label{IWg7}
k_{3} = \frac{1}{1440\pi^{2}}[N_{0} + \frac{11}{2}N_{1/2}+31 N_{1}]
\end{equation}
where $N_{0}, N_{1/2}$ and $N_{1}$ are the numbers of quantum fields with spin $0, \frac{1}{2}$ and $1$ respectively. 

Introducing
\begin{equation}
\label{IWg8}
H_{0} \equiv (8 \pi k_{3}G_{\mathrm{N}})^{-1/2}, \quad \tilde{M} \equiv (48\pi k_{1}G_{\mathrm{N}})^{-1/2}
\end{equation}
then (\ref{IWg1}) becomes, for a flat FRW metric,
\begin{equation}
\label{IWg10}
H^{2}(H^{2}-H_{0}^{2}) = \frac{H_{0}^{2}}{\tilde{M}^{2}}(2H \ddot{H}+6H^{2}\dot{H}-\dot{H}^{2}).
\end{equation}

  Cosmological gravitational production of particles~\cite{KL2024} (CGPP) is represented by a 
\textit{spectator} scalar field $\psi$ of mass $m$ which has a 
conformal coupling $\xi R \psi^{2}$ to the Ricci scalar. This cosmological particle production is 
treated as a semiclassical process in that the gravitational field is not quantized and is assumed to have a 
rigid geometry unaffected by the particle production. 
The field $\psi$ is quantized in terms of a set of mode functions $\chi_{k}(\eta)$, 
where $\eta$ is the conformal time $d\eta \equiv dt/a$ but, as the metric is non-stationary, 
a new set is required for different values of $\eta$. In terms of the sets
$\chi_{k}^{\mathrm{in}}(\eta)$ and $\chi_{k}^{\mathrm{out}}(\eta)$ valid for $\eta \rightarrow -\infty$ and $\eta \rightarrow \infty$
respectively, the comoving number density $n$ and energy density $\rho$ of the produced particles are
\begin{equation}
\label{IW900}
n=\frac{1}{(2\pi)^{3}} \frac{1}{a^{3}} \int |\beta_{k}|^{2} d^{3}k, 
\end{equation}
and
\begin{equation}
\label{IW900a}
\rho=\frac{1}{(2\pi)^{3}} \frac{1}{a^{4}} \int \omega_{k}|\beta_{k}|^{2} d^{3}k,
\end{equation}
where
\begin{equation}
\label{IW901}
\beta_{k} = -i \int^{\infty}_{-\infty} \chi_{k}^{\mathrm{in}}(\eta)\mathcal{V}(\eta)\chi_{k}^{\mathrm{out}}(\eta)  d \eta
\end{equation}
and
\begin{equation}
\label{IW902}
\mathcal{V}(\eta) =m^{2}[C(\infty)-C(\eta) +(\frac{1}{6}-\xi)C(\eta)R(\eta).
\end{equation}
Here $C(\eta)=a^{2}(\eta)$ and $\omega^{2}_{k}=k^{2}+m^{2}C(\infty)$. 

\cite{KL2024} show that the renormalized energy density for an observer with 4-velocity $U^{\mu}$
\begin{equation}
\label{IW49a}
\rho^{\mathrm{ren}}(x) = \langle 0^{\mathrm{in}}|N[T_{\mu\nu}U^{\mu}U^{\nu}]|0^{\mathrm{in}}\rangle,
\end{equation}
where $T_{\mu\nu}$ is the energy-momentum tensor calculated from the Lagrangian density of the $\psi$ field, 
and the normal ordering $N$ is performed with respect to the \textit{out} basis ladder operators,
reduces to (\ref{IW900a}) for the typical case of inefficient CGPP, where $|\beta_{k}| \ll 1$,
and for a cosmological FRW spacetime where $H(\eta)$ and $R(\eta)$ are monotonically decreasing.

The factors $a^{-3}$ and $a^{-4}$ reflect the dilution of particles by the expansion after their creation. Particle 
creation is regarded as having effectively ceased after $\mathcal{V}(\eta)$ and $d\mathcal{V}(\eta)/d \eta$ have 
become small compared to their maximum values. From that time onwards, the particles behave as a ultrarelativistic gas.

For the case $m=0$, the number density to lowest order in $\mathcal{V}(\eta)$, is
\begin{equation}
\label{IWg102}
n= \frac{1}{576 \pi a^{3}} (1-6\xi)^{2} \int C^{2}(\eta)R^{2}(\eta) d\eta. 
\end{equation}
The gravitational particle production at time $t$ is then
\begin{equation}
\label{IWg103}
n(t) =\frac{1}{576 \pi a^{3}} (1-6 \xi)^{2} \int^{t}_{-\infty} 
a^{3}(t) R^{2}(t) dt.
\end{equation}
This is valid in any $f(R)$ model. For the Starobinsky $R^{2}$ model, the energy density 
becomes~\cite{MN2012} 
\begin{equation}
\label{IWg104}
\rho(t) = \frac{g_{\ast}(T)\tilde{M}}{1152 \pi a^{4}}  (1-6\xi)^{2} 
\int^{t}_{-\infty} 
a^{4}(t) R^{2}(t) dt.
\end{equation}

Reheating due to gravitational production of particles has been studied  for the Starobinsky model~\cite{MMS1986}
and for the $gR^{2}-AB$ $f(R)$ model (\ref{IW209a})\cite{ABS2010,MN2012}. 
The  $f(R)$ gravity $(0,0)$ component of the Friedmann equation (\ref{IW103}) can be rewritten as
\begin{equation}
\label{IW553i}
6 f^{\prime\prime}(R) H(\ddot{H} + 4H\dot{H}) + \frac{f(R)}{6}
-  (\dot{H}+H^{2})f^{\prime}(R) =\frac{\kappa^{2}}{3} \rho.
\end{equation}
This is a nonlinear 4th order differential equation which is complicated to solve. 

\cite{ABS2010} first solve (\ref{IW553i}) for the inflation era where, for $\tilde{M}^{2} < R < M_{\mathrm{Pl}}^{2}$, 
$f(R)=R+R^{2}/6 \tilde{M}^{2}$ to an excellent approximation. 
Equation (\ref{IW553i}) then becomes (for $\rho=0$)
\begin{equation}
\label{IW553j}
2 H \ddot{H} + 6 H^{2}\dot{H}  - \dot{H}^{2}+ \tilde{M}^{2} H^{2} =0.
\end{equation}
Note that (\ref{IW553j}) is the $H \ll H_{0}$ limit of (\ref{IWg10}) for the quantum-corrected Einstein's equations. 

For slow-roll evolution $H^{2} \gg |\dot{H}|$, $H|\dot{H}| \gg |\ddot{H}|$, (\ref{IW553j}) reduces to
\begin{equation}
\label{IW107a}
\dot{H} = - \frac{\tilde{M}^{2}}{6},
\end{equation}
and the Hubble parameter is  
\begin{equation}
\label{IW107b}
H(t) \approx H_{i} - \frac{\tilde{M}^{2}}{6} t,
\end{equation}
where $t=0$ at the beginning of the evolution and $H_{i} \leq M_{\mathrm{Pl}}$. Slow roll ends
when $H \sim \tilde{M}$.

\cite{ABS2010} then solve (\ref{IW553i}) numerically for the epoch immediately following slow-roll, initially neglecting 
the effects of any gravitational particle production. During the oscillation (reheating) phase, it is assumed that the 
energy density (\ref{IWg104}) is totally converted to a radiation density 
\begin{equation}
\label{IWg104a}
\rho(t)=\rho_{\mathrm{rad}}=\frac{\pi^{2}}{30} g_{\ast}^{4}(T). 
\end{equation}

Motohashi and Nishizawa~\cite{MN2012} argue that the dynamics of inflation and reheating is more intuitively understood 
by studying the motion of the scalaron from its potential in the Einstein frame. They consider two regimes, 
the slow-roll inflationary era and the fast-roll oscillation era, and derive analytic solutions for $\phi(t_{\mathrm{E}})$, 
$ H_{\mathrm{E}}(t_{\mathrm{E}})$, $a_{\mathrm{E}}(t_{\mathrm{E}})$ and $R_{\mathrm{J}}(t_{\mathrm{E}})$ 
for the slow roll and fast roll approximations. The time variable $t_{\mathrm{E}}$ is then converted into $t_{\mathrm{J}}$.
Finally, $\rho_{\mathrm{J}}(t_{\mathrm{J}})$ is calculated from (\ref{IWg104}). This will be discussed in Sec. VI.

It is generally assumed\cite{MN2012} that, since the energy density of the created radiation is sub-dominant compared to the energy 
density of the inflation during both the slow-roll and oscillation regimes, 
its back reaction on the background dynamics can be neglected in obtaining analytical solutions.

\section{Existing calculations of gravitational baryogenesis}

The early calculation by Davoudiasl et al \cite{DKKMS2004} is based upon General Relativity.  
Using $R=-\kappa^{2}T =-\kappa^{2} \rho (1-3w)$, where $w$ is the equation of state parameter $p=w \rho$, 
together with the energy conservation equation and Friedmann's equation, gives
\begin{equation}
\label{IW300}
\dot{R} =-\sqrt{3} \kappa^{3} (1-3w) (1+w) \rho^{3/2}.
\end{equation}
The Baryon asymmetry factor is then
\begin{eqnarray}
\label{IW303}
\frac{\delta n_{\mathrm{b}}}{s} & = & \sqrt{3} 
\left( \frac{15 g_{\mathrm{B}}(1-3w)(1+w)}{4 \pi^{2}g_{\ast}(T)M_{\ast}^{2} M_{\mathrm{Pl}}^{3} T} \right)\rho^{3/2}.
\end{eqnarray}

Davoudiasl et al\cite{DKKMS2004} consider three scenarios to realize a non-zero $\dot{R}$, that is, with $w \neq 1/3$.

The first is a radiation-dominated era following inflation when $w \approx 1/3$. In the limit of exact 
conformal invariance, $T^{\mu}_{\mu}=0$ and $w=1/3$. However, interactions among massless particles lead to 
running coupling constants and hence the trace anomaly
\begin{equation}
\label{IW301}
T^{\mu}_{\mu} \propto \beta(g) F^{\mu\nu} F_{\mu\nu} \neq 0.
\end{equation}
Typical gauge groups and matter content at very high energies give $1-3w \sim 10^{-2} - 10^{-1}$.
In this scenario, the decoupling temperature $T_{\mathrm{D}}$ is less than the temperature of radiation dominance 
$T_{\mathrm{RD}}$ (when the decay rate of the inflaton into radiation satisfies $\Gamma \sim H$). 
The baryon asymmetry factor is
\begin{equation}
\label{IW305}
\frac{\delta n_{\mathrm{b}}}{s}  = 15\pi g_{\mathrm{B}}\sqrt{\frac{g_{\ast}(T)}{10}}
(1+w)(1-3w)\frac{T^{5}}{M_{\ast}^{2} M_{\mathrm{Pl}}^{3}}.
\end{equation}
If $M_{\ast} \sim M_{\mathrm{Pl}}$ then $\delta n_{\mathrm{b}}/s$ can be sufficiently large if 
$T_{\mathrm{D}} \sim M_{\mathrm{inf}}$, where $M_{\mathrm{inf}}$ is the inflation scale, of order $10^{16}$ GeV. In this case
\begin{equation}
\label{IW306}
\frac{\delta n_{\mathrm{b}}}{s} \approx 4.6 \times 10^{-10} (1-3w).
\end{equation}

The second scenario is a matter-dominated era $w=0$ such as reheating via a scalar field $\phi_{\mathrm{osc}}$
as it oscillates about the minimum of a quadratic potential~\cite{Turner1983}.  The field decays into radiation
at the rate $\Gamma$, with $T_{\mathrm{D}}> T_{\mathrm{RD}}$, and the final asymmetry is
\begin{equation}
\label{IW315}
\frac{\delta n_{\mathrm{b}}}{s} \leq 10^{2} \frac{T_{\mathrm{D}}^{5}}{M_{\ast}^{2} M_{\mathrm{Pl}}^{3}}.
\end{equation}
This is three to four orders of magnitude enhanced relative to (\ref{IW305}) and allows for the temperature
$T_{\mathrm{RD}}$ to be $\approx 10^{14}$ GeV.

The third scenario is to generate the asymmetry while a non-thermal component with $w > 1/3$ dominates the 
Universe. A general $w$ can be realized, for example, by a coherent oscillation of $\phi$ about the minimum of the potential
\begin{equation}
\label{IW316}
V(\phi) = \frac{\lambda \phi^{2N}}{M_{\mathrm{Pl}}^{2N-4}}.
\end{equation}
This form of potential yields~\cite{Turner1983}
\begin{equation}
\label{IW317}
w= \frac{N-1}{N+1},
\end{equation}
that is $\frac{1}{3} < w \leq 1$ for $N>2$. The field density $\rho_{\phi} \sim a^{-3(1+w)}$ initially dominates the Universe
but decreases faster than the sub-dominant radiation $\rho_{\mathrm{rad}} \sim a^{-4}$, leading to an asymmetry
\begin{equation}
\label{IW324}
\frac{\delta n_{\mathrm{b}}}{s}  =  15\pi g_{\mathrm{b}}\sqrt{\frac{g_{\ast}(T)}{10}} (1+w)(1-3w)
\frac{T^{8}}{M_{\ast}^{2} M_{\mathrm{Pl}}^{3}T_{\mathrm{RD}}^{3}} 
\left(\frac{T_{\mathrm{RD}}}{T}\right)^{9(1-w)/2}.
\end{equation}
This has to be evaluated at $T=T_{\mathrm{D}}$. This scenario allows $T_{\mathrm{D}}$ to be
significantly larger than $T_{\mathrm{RD}}$.

Since $w > 1/3$, the asymmetry (\ref{IW324}) is enhanced relative to that in (\ref{IW305}) by a factor
\begin{equation}
\label{IW325}
\left(\frac{T_{\mathrm{D}}}{T_{\mathrm{RD}}}\right)^{3(3w-1)/2},
\end{equation}
thus favoring the case $w >1/3$.

Lambiase and Scarpetta\cite{LS2006} consider baryogenesis in a $f(R)$ gravity with the power law~\cite{KKKC2015} 
\begin{equation}
\label{IW160}
f(R) = \left(\frac{R}{A}\right)^{n}.
\end{equation}
They assume a radiation dominated early Universe and apply the ansatz $a(t) \sim t^{\alpha}$, 
giving
\begin{equation}
\label{IW327}
\frac{\delta n_{\mathrm{b}}}{s} =  \frac{45}{\pi^{2}} \frac{g_{\mathrm{B}}}{g_{\ast}(T)} 
\frac{\alpha (1-2 \alpha)}{t^{3}T M_{\ast}^{2}}.
\end{equation}

Noting that the time dependence $f^{\prime}(R)/\rho \propto t^{-2}$ requires $n=2 \alpha$, 
the Friedmann equation (\ref{IW103}) becomes
\begin{eqnarray}
\label{IW164}
\rho & = & \frac{A^{-n}}{2 \kappa^{2}} \tilde{g}_{2\alpha} t^{-4\alpha}
\nonumber  \\
& = & \frac{\pi^{2}}{30} g_{\ast}(T) T^{4},
\end{eqnarray}
where
\begin{equation}
\label{IW329}
\tilde{g}_{2\alpha} \equiv (6 \alpha)^{2\alpha} \frac{[-10\alpha^{2}+8 \alpha-1]}{2(1-2\alpha)^{(1-2\alpha)}},
\end{equation}
which is positive for $0.155 \leq \alpha \leq 1/2$. The asymmetry is 
\begin{equation}
\label{IW330}
\frac{\delta n_{\mathrm{b}}}{s} =  2\times 6^{-3/2}\left(\frac{45 g_{\mathrm{b}}}{\pi^{2}g_{\ast}(T)} \right)H_{\alpha} 
\left(\frac{A}{M_{\mathrm{P}}^{2-1/\alpha}}\right)^{3/2}
\left(\frac{T}{M_{\mathrm{P}}}\right)^{3/\alpha-1}\left(\frac{M_{\mathrm{P}}}{M_{\ast}}\right)^{2}
\end{equation}
where $M_{\mathrm{P}}\equiv \sqrt{8 \pi}M_{\mathrm{Pl}}$, and 
\begin{equation}
\label{IW332}
H_{\alpha} \equiv \frac{1}{\sqrt{\alpha (1-2\alpha)}} 
\left[ \frac{4 \pi^{3}g_{\ast}(T)}{15}\,\frac{2(1-2\alpha)}{-10\alpha^{2}+8 \alpha-1}\right]^{3/4\alpha}.
\end{equation}

Assuming $M_{\ast} = M_{\mathrm{Pl}}/\sqrt{8\pi}$ for $T_{\mathrm{D}} =M_{\mathrm{inf}}$, and 
$A \sim M_{\mathrm{P}}^{2-1/\alpha}$, then the constraint $n_{\mathrm{b}}/s \leq 9 \times 10^{-11}$
and BBN require $0.46 \leq \alpha \leq 0.49$. The preferred value $\alpha = 0.4645$ corresponds to 
$n \approx 0.97$, and a function $f(R)$ which is nearly linear in $R$.

Pizza~\cite{Pizza2015} considers the Starobinsky model
\begin{equation}
\label{IW333}
f(R) = R + \alpha R^{2}
\end{equation}
and analyzes it assuming the ansatz for a slightly modified radiation dominated phase
\begin{equation}
\label{IW334}
\tilde{a}(t) = a_{0} t^{1/2} + \lambda (t),
\end{equation}
where $\lambda (t)$ has only a slight effect on the thermal history of the Universe.

The Friedmann equation (\ref{IW103}), retaining only terms linear in $\lambda$ and assuming 
$\lambda(t)=\lambda_{0}t^{\beta}$, gives
\begin{equation}
\label{IW348}
1 + 2 (2 \beta -1)\left[ 1+ 3(2 \beta +1)(\beta -2) \frac{\alpha}{t^{2}}\right] \frac{\lambda}{a} =0.
\end{equation}
Defining 
\begin{equation}
\label{IW349}
\epsilon \equiv \frac{\lambda}{a}, \quad \gamma \equiv 1-2 \beta ,
\end{equation}
then the baryon asymmetry factor (\ref{IW42}) is
\begin{equation}
\label{IW355}
\frac{\delta n_{\mathrm{b}}}{s}  =  \frac{3 \epsilon_{0}}{4} \left[\frac{15 g_{\mathrm{b}}}{4 \pi^{2}g_{\ast}(T)} \right]
\left[\frac{16 \pi^{3} g_{\ast}(T)}{45} \right]^{3/2} 
\gamma (2-\gamma)(\gamma+4) \frac{T^{\gamma+6}}{M_{\mathrm{P}}^{\gamma+3}}\frac{1}{M_{\ast}^{2}T}
\end{equation}
where
\begin{equation}
\label{IW352}
\epsilon = \epsilon_{0} \left(\frac{T}{M_{\mathrm{P}}}\right)^{\gamma}
\end{equation}
and
\begin{equation}
\label{IW353}
\epsilon_{0} = \frac{\lambda_{0}}{a_{0}} \left[\frac{4 \pi}{3} \sqrt{\frac{\pi g_{\ast}(T)}{5}} \right]^{\gamma/2}
M_{\mathrm{P}}^{\gamma/2}.
\end{equation}

Choosing $M_{\ast} = M_{\mathrm{P}}$ then
\begin{equation}
\label{IW356}
\frac{\delta n_{\mathrm{b}}}{s} = \frac{3 \epsilon_{0}}{4} A_{1} A_{2} \gamma (2-\gamma)(\gamma+4) 
\frac{T^{\gamma+5}}{M_{\mathrm{P}}^{\gamma+5}}
\end{equation}
where, using $g_{\mathrm{B}} =2$ and $g_{\ast}(T)=106.75$,
\begin{equation}
\label{IW357}
A_{1} \equiv \left[\frac{15 g_{\mathrm{B}}}{4 \pi^{2}g_{\ast}(T)} \right] = 7.1186 \times 10^{-3},
\end{equation}
and
\begin{equation}
\label{IW358}
A_{2} \equiv \left[\frac{16 \pi^{3} g_{\ast}(T)}{45} \right]^{3/2} = 4.0373 \times 10^{4}.
\end{equation}
Note that the author\cite{Pizza2015} does not include the factor $A_{1}$ in their expression for
the baryon asymmetry.

Choosing $T_{\mathrm{D}}= 10^{16}$ GeV, that is $(T_{\mathrm{D}}/M_{\mathrm{P}})^{5} = 3.6864 \times 10^{-16}$,
and choosing the arbitrary ratio $\lambda_{0}/a_{0}$ small enough to make $\epsilon_{0}=1$, then
$\eta \equiv (\delta n_{\mathrm{b}}/s)/A_{1}$ is equal to $4.665 \times 10^{-12}$, $4.456 \times 10^{-12}$, 
$2.010 \times 10^{-12}$ and $5.070 \times 10^{-14}$ for $\gamma =(0.1, 0.2, 0.4, 1.0)$. Thus the Pizza model
is unable to generate the observed asymmetry of $\approx 10^{-10}$.

More recently, Ramos and P\'{a}ramos~\cite{RP2017} have generalized gravitational baryogenesis to nonminimally 
coupled $f(R)$ gravity where the action (\ref{IW48}) is extended to 
\begin{equation}
\label{IW48a}
\mathcal S  = \int d^{4}x \sqrt{-g} [\frac{1}{2 \kappa^{2}}f_{1}(R) +f_{2}(R)\mathcal L_{m}(g_{\mu \nu},\psi_{m})].
\end{equation}
They make the same ansatz $a(t) \sim t^{\alpha}$ as Lambiase and Scarpetta\cite{LS2006}, and choose the power-law forms
\begin{equation}
\label{IW48b}
f_{1}(R)=R\left(\frac{|R|}{M_{1}^{2}}\right)^{m}, \quad f_{2}(R)=R\left(\frac{|R|}{M_{2}^{2}}\right)^{n}.
\end{equation}
The baryon asymmetry is again given by (\ref{IW327}), and, for the choice $M_{\ast} \sim M_{\mathrm{Pl}}$, is 
sufficiently large if $T_{\mathrm{D}} = M_{\mathrm{inf}} \approx 2 \times 10^{16}$ GeV. 

They study numerically the constraints on the exponents $(m,n)$ and mass scales $M_{1,2}$ for different choices of the 
decoupling temperature $T_{\mathrm{D}}$.

As a special case, they consider the minimal coupling $n=0$ case. This is a generalization of the Lambiase and Scarpetta
calculations for the exponent $m$ and mass scale $M_{1}$. They conclude that $-0.07 \simeq m \simeq 0.19$, 
for $T_{\mathrm{D}} = 2 \times 10^{16}$ GeV, but that $M_{1} \sim M_{\mathrm{Pl}}$ and could affect inflation.
They therefore study the modified Starobinsky form
\begin{equation}
\label{IW48c}
f_{1}(R) =R\left(\frac{|R|}{M_{1}^{2}}\right)^{m} + \frac{R^{2}}{6 M_{\mathrm{S}}}
\end{equation}
and find that, for $T_{\mathrm{D}} = 2 \times 10^{16}$ GeV, the mass scales are much smaller, $M_{1} \sim 10^{12}$ GeV.
They are strongly dependent on $T_{\mathrm{D}}$ for small values of $m$, with $M_{1} \sim T_{\mathrm{D}}^{-5/3m}$
for $m \sim 0$.

Gravitational baryogenesis has been studied in several other modified gravities.  We will briefly outline some of these studies.

Odintsov and Oikonomou consider Gauss-Bonnet baryogenesis~\cite{OO2016} based on the coupling 
$j^{\mu}_{\mathrm{B}} \partial_{\mu}\mathcal{G}$, where 
$\mathcal{G} =R^{2}-4R_{\mu\nu}R^{\mu\nu}+R_{\mu\nu\rho\sigma}R^{\mu\nu\rho\sigma}$ 
is the Gauss-Bonnet curvature invariant, for two gravity models. 
The first model is a Universe described by the standard Einstein-Hilbert gravity, filled with a 
perfect fluid with a constant $w=p/\rho$, and with a flat FRW metric and assumed to evolve as $a(t) = B t^{\beta}$,
where $\beta=2/(3(1+w))$. The asymmetry has the dependence $\sim T_\mathrm{D}^{5/\beta-1}$, and is non-zero for the radiation 
domination case $\beta=1/2$. The second model is an evolution determined by an $R+f(\mathcal{G})$ gravity for a matter field
with constant $w=p_{m}/\rho_{m}$, where it is assumed that $f(\mathcal{G})=f_{0}\mathcal{G}^{\gamma}$. 
For $\gamma <1/2$, the evolution is $a(t) = B t^{\beta}$ with $\beta=4\gamma/(3(1+w))$. The asymmetry has the dependence 
$\sim T_\mathrm{D}^{5/\gamma-1}$, and depends on the parameters $\gamma$ and $f_{0}$, that can be chosen to meet the 
observational bound for baryon asymmetry. 

Odintsov and Oikonomou also consider~\cite{OO2016a} the effects of Loop Quantum Cosmology (LQC) on the interaction  
$j^{\mu}_{\mathrm{B}} \partial_{\mu}R$, for an Einstein-Hilbert Universe with a flat FRW metric. These effects produce an 
additional term $\propto (1+3w)\rho^{2}/\rho_{c}$, where the parameter $\rho_{c}$ represents the quantum effects of the theory, 
in the General Relativity expression (\ref{IW300}) for $\dot{R}$.  The asymmetry depends on $w$, is non-zero for the 
radiation domination $w=1/3$ case, and is compatible with observations for a range of values of $w \leq 1$, indicating that LQC 
strongly affects the baryon asymmetry.

Oikonomou and Saridakis~\cite{OS2016} have studied baryogenesis in $f(T)$ gravity, where $R$ is replaced by the torsion scalar $T$,
and the asymmetry depends on $\dot{T}$. The effects of couplings to $\partial_{\mu}T$ and $\partial_{\mu}f(T)$ on the baryon asymmetry
for a flat FRM metric (in which $T=-6H^{2}$) are examined for the cases $f(T)=T$ and $f(T) \sim (-T)^{n}$. 
The asymmetry is, again, non-zero for a radiation-dominated Universe. 
The first case is not compatible with the observed asymmetry for acceptable parameter choices but the second case, 
where the asymmetry depends crucially on $n$, can produce compatible values for the coupling to $\partial_{\mu}f(T)$.

As alternatives to these models, in which the baryon asymmetry arises from gravitational corrections to the Einstein-Hilbert action,
Pereira et al have studied baryogenesis generated by higher-order matter contributions incorporated in a 
$f(R,\mathcal{T}^{2})$ gravity, where $\mathcal{T}^{2}\equiv T_{\mu\nu}T^{\mu\nu}$, and by gravitational scalars $\phi_{i}$ in a 
scalar-tensor representation of modified gravity. 

The first study~\cite{PLM2025} assumes the gravitational model 
$f(R,\mathcal{T}^{2})=R+\eta M_{\mathrm{Pl}}^{2-8n} (\mathcal{T}^{2})^{n}$, where $n=1/2$ and $n=1$, and the interaction
$\sim \partial_{\mu}(\mathcal{T}^{2})j_{\mathrm{B}}^{\mu}$. The asymmetry is $\sim \dot{\mathcal{T}}^{2}$. 
For a perfect fluid, $\mathcal{T}^{2}=\rho^{2}(1+3w)$, and the interaction effectively couples $j_{\mathrm{B}}^{\mu}$ 
with $\partial_{\mu}(\rho^{2})$, suggesting high energy matter can enhance baryon asymmetry.

Scalar-tensor baryogenesis~\cite{Pereira2026} is based upon the derivative coupling 
$M_{\ast}^{d}\nabla_{\mu} f(\phi_{i}) j_{\mathrm{B-L}}^{\mu}$, where $f(\phi_{i})$ is a general function of the scalars $\phi_{i}$.
The theory is illustrated for $f(R)$ gravity in which its scalar field is $\phi=df/dR$, the interaction is 
$\sim \partial_{\mu}\phi j^{\mu}$, and the asymmetry is $\sim \dot{\phi}$. The case $f(R)=M_{\mathrm{Pl}}^{2-2\epsilon}R^{1+\epsilon}$
is studied and viable baryogenesis obtained for $\epsilon =O(10^{-6})$ with $T_{\mathrm{D}}=8.5 \times 10^{13}$ GeV.

\section{Implications of recent measurements of the CMB} 

Measurements of the primordial scalar spectrum of the CMB provide precision tests of single-field inflationary models 
within the framework of Einstein's General Relativity~\cite{Planck2018a}. The observational data favour slow-roll models 
with concave scalar potentials. In terms of the slow-roll parameters  
\begin{equation}
\label{IW927}
\epsilon \equiv \frac{M_{\mathrm{Pl}}^{2}}{2} \left(\frac{U_{\phi}}{U}\right)^{2}, \quad
\eta \equiv  M_{\mathrm{Pl}}^{2} \frac{U_{\phi\phi}}{U}, 
\end{equation}
where, eg, $U_{\phi} \equiv dU/d\phi$, measurements of the scalar tilt 
\begin{equation}
\label{IW928}
n_{s} \simeq 1-6 \epsilon_{\ast} + 2 \eta_{\ast},
\end{equation}
where the subscript $\ast$  denotes a suitable pivot scale, and the scalar-to-tensor ratio
\begin{equation}
\label{IW928a}
r \simeq 16 \epsilon_{\ast},
\end{equation}
constrain models of inflation.

The experimental value of $n_{s}$ at the pivotal scale $k_{\ast}=0.05$ Mpc$^{-1}$ extracted from the 
Planck 2018 data \cite{Planck2018a,Planck2018b}, in combination with gravitational lensing, is
\begin{equation}
\label{IW929}
n_{s} = 0.9649 \pm 0.0042 \,(68\%\,\mathrm{CL}).
\end{equation}
This becomes $n_{s}=0.9652\pm 0.0042$ when combined with BICEP/Keck data ~\cite{BICEP2021}.  The current 95\% CL 
upper bound on $r$ is $0.036$.

The Starobinsky model gives $n_{s}=0.965$ for $N_{\ast}=55$ and $r=0.0035$, consistent with the value (\ref{IW929}).
$N_{\ast}$ is the number of e-folds starting from the pivot scale:
\begin{equation}
\label{IW928b}
N_{\ast} \equiv \log \left(\frac{a_{\mathrm{end}}}{a_{\ast}}\right),
\end{equation}
where $a_{\mathrm{end}}$ is the cosmological scale factor at the end of inflation. 
However, this requires the low value $M/M_{\mathrm{P}} \simeq 1.3 \times 10^{-5}$ for $N_{\ast}\simeq 55$, and the large 
initial value $\phi \simeq 5.5$. Note that $M=\sqrt{6}\tilde{M}$.

Recently two new high-multipole CMB measurements have been reported by the Atacama Cosmology Telescope (ACT) 
collaboration~\cite{Louis2025,Calabrese2025}  and the South Pole Telescope (SPT)-3G collaboration~\cite{Camphuis2025}. 
The ACT measurement $n_{s}=0.9666\pm 0.0077$ at the 68\% CL is compatible
with (\ref{IW929}) but a fit to Planck, lensing and ACT yields a higher value $n_{s}=0.9713 \pm 0.0037$ at 68\% CL, in some
tension with the Planck and lensing combination (\ref{IW929}). This tension is further increased if Dark Energy 
Spectroscopic Instrument (DESI) measurements of baryon acoustic oscillations (BAO) are combined with the 
Planck, lensing and ACT data (P-ACT-LB):
\begin{equation}
\label{IW930}
n_{s} = 0.9743 \pm 0.0034 (68\%\,\mathrm{CL}).
\end{equation}

The SPT-3G measurements give $n_{s}=0.951 \pm 0.011$ at 68\% CL value (\ref{IW929}), below the Planck/lensing 
value (\ref{IW929}). Combining Planck and ACT data with SPT data gives
\begin{equation}
\label{IW931}
n_{s} = 0.9684 \pm 0.0030 (68\%\,\mathrm{CL}).
\end{equation}

The effect of these recent measurements on $f(R)$ gravity models of inflation has been investigated by 
Ketov et al~\cite{Ketov2025}. They
undertake a detailed study of the evolution equations in the Jordan frame and propose a new description of the 
slow-roll approximation. They argue that the Starobinsky description of inflation provides merely the 
leading (or dominant) terms in the gravitational effective action and in the inflaton potential. 
Higher-order terms will be important to account for more precise CMB measurements in the future.

They consider the key CMB observables: the scalar tilt $n_{s}$ and the running index $\alpha_{s}\equiv dn_{s}/d\,\log k$,
for which the Planck/BICEP data are
\begin{equation}
\label{IW935}
n_{s}=0.9651 \pm 0.0044, \quad \alpha_{s} =-0.0069 \pm 0.0069,
\end{equation}
and the more recent ACT/DESI data are
\begin{equation}
\label{IW936}
n_{s}=0.9743 \pm 0.0034, \quad \alpha_{s} =0.0062 \pm 0.0052.
\end{equation}

\cite{Ketov2025} comment that it is possible to increase the value of $n_{s} $ in the Starobinsky model by 
increasing the number of e-folds beyond 70 but
that is unacceptable, and that it is impossible to get a positive running index $\alpha_{s}$.

The small deformations $\delta_{3}R^{3}/36 \tilde{M}^{4}$ or $\delta_{4}R^{4}/48 \tilde{M}^{6}$ 
of Starobinsky inflation can give values of $n_{s}$ in agreement with the ACT results, but the values 
of $\alpha_{s}$ remain negative. They propose the new model
\begin{equation}
\label{IW939}
f_{5}(R)  =  R + \frac{R^{2}}{6 \tilde{M}^{2}} + c_{3}\frac{R^{3}}{\tilde{M}^{4}}
+ c_{4}\frac{R^{4}}{\tilde{M}^{6}} + c_{5}\frac{R^{5}}{\tilde{M}^{8}},
\end{equation}
where $c_{3},c_{4}$ and $c_{5}$ are dimensionless coupling constants. They find that the values of $c_{i}$ can be chosen
to meet the ACT favoured values of $n_{s}$ and $\alpha_{s}$. 

Odintsov and Oikonomou~\cite{OO2025} have revisited power-law $f(R)$ gravity in light of the ACT observations.
In their new model, \cite{OO2025} commence with the slow-roll indices for $f(R)$ gravity:
\begin{equation}
\label{IW950}
\epsilon_{1} = -\frac{\dot{H}}{H^{2}}, \quad \epsilon_{3}=\frac{\dot{f}^{\prime}}{2H f^{\prime}}, \quad
\epsilon_{4} = \frac{\ddot{f}^{\prime}}{H \dot{f}^{\prime}}.
\end{equation}
Assuming these indices $\epsilon_{i} \ll 1$, then
\begin{equation}
\label{IW951}
n_{s}=1-4\epsilon_{1} +2\epsilon_{3} -2 \epsilon_{4}, \quad r=48 \frac{\epsilon_{3}^{2}}{(1+\epsilon_{3})^{2}}.
\end{equation}

The Raychaudhuri equation for $f(R)$ gravity gives 
\begin{equation}
\label{IW952}
\epsilon_{1}=-\epsilon_{3}(1-\epsilon_{4}) \simeq - \epsilon_{3}
\end{equation}
so that
\begin{equation}
\label{IW953}
n_{s} \simeq 1-6 \epsilon_{1} - 2 \epsilon_{4}, \quad r \simeq 48 \epsilon_{1}^{2}.
\end{equation}
Introducing the dimensionless parameter
\begin{equation}
\label{IW954}
x \equiv \frac{48 f^{\prime\prime\prime}H^{2}}{f^{\prime\prime}}
\end{equation}
then, under slow-roll conditions, 
\begin{equation}
\label{IW954a}
\epsilon_{4} \approx -\frac{x}{2} \epsilon_{1} - \epsilon_{1}
\end{equation}
and
\begin{equation}
\label{IW955}
n_{s} = 1-4 \epsilon_{1} + x \epsilon_{1}, \quad r =\frac{48(1-n_{s})^{2}}{(4-x)^{2}}.
\end{equation}
Hence, in this formalism, it is necessary to calculate $x$ and $\epsilon_{1}$ to obtain the inflationary phenomenology.

Using the slow-roll inflationary approximation $R \sim 12 H^{2}$ then
\begin{equation}
\label{IW956}
x = \frac{4 f^{\prime\prime\prime}R}{f^{\prime\prime}}.
\end{equation}
If $x$ is constant, say $x=k$, then (\ref{IW956}) can be solved analytically, giving the exact power law evolution
\begin{equation}
\label{IW961}
f(R) = \frac{16 c_{1}}{(4+k)(8+k)} R^{2+k/4} + c_{2}R + c_{3},
\end{equation}
where the $c_{i}$ are dimensionful integration constants.

This gives
\begin{equation}
\label{IW962}
\epsilon_{1} = - \frac{2k}{k^{2}+6k+8} + \frac{2 c_{2}R^{-1-k/4}}{c_{1}(k+2)}.
\end{equation}

The model is compatible with the Planck data for $k$ in the range $[-0.038,-0.03]$ and with ACT for $k=[-0.0282,-0.022]$.
 
For $k=0$, (\ref{IW961}) becomes
\begin{equation}
\label{IW961a}
f(R)=\frac{c_{1}}{2} R^{2} +c_{2}R + c_{3},
\end{equation}
which is the Starobinsky form for $c_{1}=1/3\tilde{M}^{2}$, $c_{2}=1$ and $c_{3}=0$.

\section{Scalaron calculations of gravitational baryogenesis during inflationary era}

\subsection{Preamble}

The calculations of baryon asymmetry (\ref{IW42}) assume the interaction $\mathcal{S}_{\mathrm{GB}}$ is treated as a
perturbation and does not affect the background dynamics~\cite{PFM2026}, that is, $R_{\mathrm{J}}(t_{\mathrm{J}})$ and 
$\dot{R}_{\mathrm{J}}(t_{\mathrm{J}})$ are calculated for an unperturbed scalaron cosmology, 
the so-called \emph{spectator approximation}.

Scalaron calculations require knowledge of the scalar potential (\ref{IW192}). However, this potential requires the relationship
$\Omega^{2}=f(R_{\mathrm{J}})$ to be inverted to obtain $R(\Omega)$. This is possible for the Starobinsky model and, as will be shown in Section 6.3, for the new power-law model of Odintsov and Oikonomou.  
Inversion of the $f_{5}(R_{\mathrm{J}})$ function (\ref{IW939}) proposed by Ketov et al is not possible.

The scalaron evolves~\cite{MMS1986} in the potential $U(\phi)$, starting from $\phi >0$ under slow roll conditions 
$H_{\mathrm{J}}^{2} \sim |\dot{H}_{\mathrm{J}}| \sim \tilde{M}^{2}$, where $R_{\mathrm{J}} \sim \tilde{M}^{2}$, 
accelerating to fast roll as it approaches $\phi=0$ where it undergoes damped harmonic oscillations around $R_{\mathrm{J}} =0$
with frequency $\omega =\tilde{M}$ and $|R_{\mathrm{J}}| \ll \tilde{M}^{2}$.

Gravitational particle production is $\sim R_{\mathrm{J}}^{2}$ and is significantly suppressed during the fast oscillation 
reheating phase where $R_{\mathrm{J}} \sim 0$. Hence, particle production effectively ceases at the end of the slow roll inflation era.

\subsection{Starobinsky model}

The dynamics of inflation and reheating has been investigated for the $gR^{2}-AB$ model by Motohashi and Nishizawa~\cite{MN2012} 
by studying the motion of the scalaron from its potential in the Einstein frame. They numerically solve the equations 
(\ref{IW83e}), (\ref{IW83j}) and the evolution equation for the energy density obtained from (\ref{IWg104}):
\begin{equation}
\label{IWg104a1}
\frac{d\rho_{\mathrm{J}}(t_{\mathrm{J}})}{dt_{\mathrm{J}}}  = -4 H_{\mathrm{J}}\rho_{\mathrm{J}} +
\frac{g_{\ast}(T)\tilde{M}}{1152 \pi a^{4}_{\mathrm{J}}}  (1-6\xi)^{2} R_{\mathrm{J}}^{2}(t_{\mathrm{J}}).
\end{equation}
This equation is in the Jordan frame and has to be translated  
to the Einstein frame using (\ref{IW122c}) and (\ref{IW122h}). After solving the three coupled equations, the
physical quantities in the Jordan frame are obtained using the inverse conformal transformations.

They also develop analytic approximate solutions for the inflationary and oscillation regimes using the 
slow roll and the fast roll  approximations, and by neglecting the back reaction of the created particles. 
Only the slow roll region is relevant in this study of gravitational baryogenesis.

The derivation of the analytic approximate solutions for the slow-roll era will be outlined here as it provides a framework
for the following treatment of the new power-law model.

For the slow roll phase, the kinetic energy is small compared to the height of the potential and the 
energy density of radiation can be neglected. The field equations (\ref{IW103e}), (\ref{IW83e}) and (\ref{IW83j}) become
\begin{eqnarray}
H_{\mathrm{E}} & = & \kappa \sqrt{\frac{U(\phi)}{3}},
\label{IW800}  \\
\frac{d H_{\mathrm{E}}}{d t_{\mathrm{E}}} & = & -\frac{\kappa^{2}}{2} \left(\frac{d \phi}{d t_{\mathrm{E}}}\right)^{2},
\label{IW801}  \\
\frac{d \phi}{dt_{\mathrm{E}}} & = & - \frac{1}{3H_{\mathrm{E}}} \frac{d U}{d \phi}
= - \frac{dU/d\phi}{\kappa \sqrt{3 U(\phi)}}.
\label{IW802}
\end{eqnarray}

In the inflation era where $f(R_{\mathrm{J}})$ has the Starobinsky form, the scalaron potential is (\ref{IW199}) and
\begin{equation}
\label{IW804}
\frac{d U}{d \phi} =\sqrt{2}\tilde{M} e^{-\sqrt{2/3}\kappa \phi}\sqrt{U(\phi)}.
\end{equation}
so that (\ref{IW802}) becomes
\begin{equation}
\label{IW805}
\frac{dt_{\mathrm{E}}}{d \phi}  =  -\frac{\kappa}{\tilde{M}}\sqrt{\frac{3}{2}}  e^{\sqrt{2/3}\kappa \phi},
\end{equation}
with the solution
\begin{equation}
\label{IW806}
t_{\mathrm{E}} -t_{\mathrm{E,init}} = - \frac{3}{2\tilde{M}} \left[e^{\sqrt{2/3}\kappa \phi} - 
e^{\sqrt{2/3}\kappa \phi_{\mathrm{init}}} \right].
\end{equation}
Assuming that the scalaron starts to roll down the potential at $t_{\mathrm{E}}=t_{\mathrm{E,init}}$ 
($t_{\mathrm{J}}=t_{\mathrm{J,init}}$), and defining
\begin{equation}
\label{IW807}
d_{\mathrm{init}} \equiv e^{\kappa \phi_{\mathrm{init}}/\sqrt{6}},
\end{equation}
then the scalaron field is
\begin{equation}
\label{IW808}
\phi(t_{\mathrm{E}}) = \frac{1}{\kappa} \sqrt{\frac{3}{2}} \ln [d_{\mathrm{init}}^{2} - 
\frac{2 \tilde{M}}{3}(t_{\mathrm{E}} -t_{\mathrm{E,init}})].
\end{equation}

The field equation (\ref{IW801}) is then
\begin{eqnarray}
\label{IW810}
\frac{d H_{\mathrm{E}}}{d t_{\mathrm{E}}} & = & -\frac{\kappa^{2}}{2} (\frac{d \phi}{d t_{\mathrm{E}}})^{2}
\nonumber  \\
& = & -\frac{\tilde{M}^{2}}{3}[d_{\mathrm{init}}^{2} - \frac{2\tilde{M}}{3}(t_{\mathrm{E}} -t_{\mathrm{E,init}})]^{-2}.
\end{eqnarray}
Integrating gives
\begin{equation}
\label{IW811}
H_{\mathrm{E}}(t_{\mathrm{E}})= \frac{\tilde{M}}{2} \left[1 - 
\frac{1}{[d_{\mathrm{init}}^{2} - \frac{2\tilde{M}}{3}(t_{\mathrm{E}} -t_{\mathrm{E,init}})]}\right].
\end{equation}
Since $H_{\mathrm{E}}(t_{\mathrm{E}}) =d (\ln a_{\mathrm{E}})/d t_{\mathrm{E}}$, integrating again gives
\begin{equation}
\label{IW815}
a_{\mathrm{E}}(t_{\mathrm{E}})  =  a_{\mathrm{E}}(t_{\mathrm{E,init}})e^{(t_{\mathrm{E}} -t_{\mathrm{E,init}})\tilde{M}/2}
\left[1 - \frac{2\tilde{M}}{3 d_{\mathrm{init}}^{2}}(t_{\mathrm{E}} -t_{\mathrm{E,init}})\right]^{3/4}.
\end{equation}

Finally, recalling (\ref{IW185}) and (\ref{IW189}), that is,
\begin{equation}
\label{IW816}
\frac{d f(R_{\mathrm{J}})}{d R_{\mathrm{J}}} = \Omega^{2} = e^{\sqrt{2/3}\kappa \phi}
\end{equation}
and using, for the Starobinsky model,
\begin{equation}
\label{IW817}
\frac{d f(R_{\mathrm{J}})}{d R_{\mathrm{J}}} = 1 + \frac{R_{\mathrm{J}}}{3 \tilde{M}^{2}}, 
\end{equation}
gives\begin{eqnarray}
\label{IW818}
R_{\mathrm{J}}(t_{\mathrm{E}}) & =  & 3\tilde{M}^{2} \left[ \Omega^{2}(t_{\mathrm{E}}) - 1\right]
\nonumber  \\
& = & 3\tilde{M}^{2} \left(e^{\sqrt{2/3}\kappa \phi(t_{\mathrm{E}})} -1\right).
\end{eqnarray}

The above results can be converted to functions of $t_{\mathrm{J}}$ using 
\begin{eqnarray}
\label{IW819}
\frac{d t_{\mathrm{J}}}{d t_{\mathrm{E}}}&  = & \Omega^{-1} (t_{\mathrm{E}})
\nonumber  \\
& = & e^{-\kappa \phi(t_{\mathrm{E}})/\sqrt{6}}
\nonumber  \\
& = & 
[d_{\mathrm{init}}^{2} - \frac{2\tilde{M}}{3}(t_{\mathrm{E}} -t_{\mathrm{E,init}})]^{-1/2},
\end{eqnarray}
that is
\begin{equation}
\label{IW820}
t_{\mathrm{J}} - t_{\mathrm{J,init}} = \frac{3}{\tilde{M}} \left[d_{\mathrm{init}} -
\sqrt{d_{\mathrm{init}}^{2} - \frac{2\tilde{M}}{3}(t_{\mathrm{E}} -t_{\mathrm{E,init}})}\right].
\end{equation}

Inverting (\ref{IW820}) gives
\begin{equation}
\label{IW821}
t_{\mathrm{E}} - t_{\mathrm{E,init}} =  (t_{\mathrm{J}} -t_{\mathrm{J,init}})   
 \left[d_{\mathrm{init}} - \frac{\tilde{M}}{6}(t_{\mathrm{J}} -t_{\mathrm{J,init}})\right].
\end{equation}

Noting that, from (\ref{IW821}),
\begin{equation}
\label{IW802a}
\left[d_{\mathrm{init}}^{2} - \frac{2\tilde{M}}{3}(t_{\mathrm{E}} - t_{\mathrm{E,init}})\right]^{1/2}
= \left[d_{\mathrm{init}} - \frac{\tilde{M}}{3}(t_{\mathrm{J}} - t_{\mathrm{J,init}})\right],
\end{equation}
then, defining
\begin{equation}
\label{IW820a}
\tau(t_{\mathrm{E}}) \equiv [d_{\mathrm{init}}^{2} - \frac{2\tilde{M}}{3}(t_{\mathrm{E}} -t_{\mathrm{E,init}})]^{1/2},
\end{equation}
and
\begin{equation}
\label{IW820b}
\tau(t_{\mathrm{J}}) \equiv \left[d_{\mathrm{init}} - \frac{\tilde{M}}{3}(t_{\mathrm{J}} - t_{\mathrm{J,init}})\right]
\end{equation}
gives
\begin{equation}
\label{IW825}
\Omega(t_{\mathrm{E}}) = \tau(t_{\mathrm{E}}) =\tau(t_{\mathrm{J}}) =\Omega(t_{\mathrm{J}}). 
\end{equation}

Conversion to functions of $t_{\mathrm{J}}$ can be affected generally by merely replacing $\tau(t_{\mathrm{E}})$ with
$\tau(t_{\mathrm{J}})$. For the Hubble parameter,
\begin{eqnarray}
\label{IW825c}
H_{\mathrm{J}}(t_{\mathrm{E}}) & = & \Omega(t_{\mathrm{E}})\left[H_{\mathrm{E}}(t_{\mathrm{E}})- \frac{\kappa}{\sqrt{6}} 
\frac{d \phi}{d t_{\mathrm{E}}}\right]
\nonumber  \\
& = & \tau(t_{\mathrm{E}}) \frac{\tilde{M}}{2} [1-\frac{1}{3} \tau(t_{\mathrm{E}})^{-2}],
\end{eqnarray}
so that
\begin{equation}
\label{IW825d}
H_{\mathrm{J}}(t_{\mathrm{J}}) = \tau(t_{\mathrm{J}}) \frac{\tilde{M}}{2} [1-\frac{1}{3} \tau(t_{\mathrm{J}})^{-2}].
\end{equation}

Recalling 
\begin{equation}
\label{IW822}
a_{\mathrm{J}}(t_{\mathrm{J}}) = \Omega^{-1} a_{\mathrm{E}}(t_{\mathrm{J}}),
\end{equation}
and changing the variable of integration from $t^{\prime}_{\mathrm{J}}$ to $t^{\prime}_{\mathrm{E}}$, the energy density of the created particles (\ref{IWg104}) is, using (\ref{IW818}), 
\begin{eqnarray}
\label{IW823}
\rho_{\mathrm{J}}(t_{\mathrm{J}})&  = & \frac{g_{\ast}(T)\tilde{M}}{1152 \pi a_{\mathrm{J}}(t_{\mathrm{J}})^{4}} (1-6\xi)^{2} 
\int^{t_{\mathrm{J}}}_{-\infty} 
a_{\mathrm{J}}^{4}(t^{\prime}_{\mathrm{J}}) R_{\mathrm{J}}^{2}(t^{\prime}_{\mathrm{J}}) dt^{\prime}_{\mathrm{J}}
\nonumber  \\
& = & \frac{9 g_{\ast}(T)\tilde{M}^{5}}{1152 \pi a_{\mathrm{E}}(t_{\mathrm{J}})^{4}} (1-6\xi)^{2} \Omega^{4}(t_{\mathrm{J}})
\nonumber  \\
& & \times \int^{t_{\mathrm{J}}}_{t_{\mathrm{J,init}}} 
\left[\Omega^{-1} (t^{\prime}_{\mathrm{E}})a_{\mathrm{E}}(t^{\prime}_{\mathrm{E}})\right]^{4}
\left[\Omega^{2}(t^{\prime}_{\mathrm{E}}) -1\right]^{2}   
\Omega^{-1} (t^{\prime}_{\mathrm{E}}) dt^{\prime}_{\mathrm{E}}.
\end{eqnarray}

Noting that, from (\ref{IW815}),
\begin{equation}
\label{IW842}
a_{\mathrm{E}}^{4}(t_{\mathrm{E}}) = a_{\mathrm{E}}^{4}(t_{\mathrm{E,init}}) 
e^{3[d_{\mathrm{init}}^{2} - \tau^{2}(t_{\mathrm{E}})]}
\left[\frac{\tau^{2}(t_{\mathrm{E}})}{d_{\mathrm{init}}^{2}}\right]^{3},
\end{equation}
and changing the integration variable from $t^{\prime}_{\mathrm{E}}$ to $\tau^{\prime}$ using 
$dt^{\prime}_{\mathrm{E}}/d \tau^{\prime} =-3/\tilde{M} \tau^{\prime}$, (\ref{IW823}) becomes
\begin{eqnarray}
\label{IW827}
\rho_{\mathrm{J}}(t_{\mathrm{J}}) & = & - \frac{3}{128 \pi} \frac{e^{3d_{\mathrm{init}}^{2}}}{d_{\mathrm{init}}^{6}}
g_{\ast}(T)\tilde{M}^{4} \left[\frac{a_{\mathrm{E}}(t_{\mathrm{E,init}})}{a_{\mathrm{E}}(t_{\mathrm{J}})}\right]^{4} 
(1-6\xi)^{2}
\nonumber  \\
& & \times \Omega^{4}(t_{\mathrm{J}})  \int^{\tau(t_{\mathrm{J}})}_{\tau_{\mathrm{init}}} 
e^{-3 \tau^{\prime\,2}} \tau^{\prime\,2} (\tau^{\prime\,2} -1)^{2} d \tau^{\prime},
\end{eqnarray}
where $\tau_{\mathrm{init}}=d_{\mathrm{init}}$.

The integral
\begin{equation}
\label{IW828}
I_{\tau}  \equiv  \int^{\tau}_{\tau_{\mathrm{init}}} e^{-3 \tau^{\prime\,2}} \tau^{\prime\,2} (\tau^{\prime\,2} -1)^{2} d \tau^{\prime}
\end{equation}
can be evaluated analytically. The result is
\begin{eqnarray}
\label{IW835}
I_{\tau} & = & - \frac{1}{432}\left\{6 \tau e^{-3\tau^{2}}[12 \tau^{4}-14 \tau^{2}+5]\right.
- 6 d_{\mathrm{init}} e^{-3 d_{\mathrm{init}}^{2}} [12 d_{\mathrm{init}}^{4} -14 d_{\mathrm{init}}^{2} +5]
\nonumber  \\
& &  \left. + 5 \sqrt{3\pi} [\mathrm{erf}(\sqrt{3}d_{\mathrm{init}}) -  \mathrm{erf}(\sqrt{3}\tau) ]\right\}.
\end{eqnarray}

Noting that $\Omega(t_{\mathrm{J}}) = \tau(t_{\mathrm{J}})$ and using (\ref{IW842}) with $t_{\mathrm{E}}$ 
replaced by $t_{\mathrm{J}}$, then the energy density of the gravitationally produced particles is 
\begin{equation}
\label{IW845}
\rho_{\mathrm{J}}(t_{\mathrm{J}}) = - \frac{3}{128 \pi} (1-6\xi)^{2} g_{\ast}(T)\tilde{M}^{4}
\frac{e^{3\tau^{2}}I_{\tau}}{\tau^{2}}. 
\end{equation}
Note that  $I_{\tau}$ is negative. The authors\cite{MN2012} use the variable $S=-\sqrt{3} \tau$ and express 
the energy density in terms of the function 
\begin{equation}
\label{IW846}
I(S) = - 1296 e^{3\tau^{2}}\,I_{\tau}.
\end{equation}

As expected, from (\ref{IW845}),
\begin{equation}
\label{IW846a}
\rho_{\mathrm{J}}(t_{\mathrm{J,init}}) = 0.
\end{equation}

The end of the slow-roll regime $t_{\mathrm{E,SR}}$ can be estimated  from when the scalaron reaches 
$\phi(t_{\mathrm{E}}) =0$ for the first time. From (\ref{IW808}), 
\begin{equation}
\label{IW847}
\phi_{\mathrm{E}}(t_{\mathrm{E,SR}}) = \frac{1}{\kappa}\sqrt{\frac{3}{2}}
\ln [d_{\mathrm{init}}^{2} - \frac{2\tilde{M}}{3}(t_{\mathrm{E,SR}} - t_{\mathrm{E,init}})] =0,
\end{equation}
that is
\begin{equation}
\label{IW848}
t_{\mathrm{E,SR}} - t_{\mathrm{E,init}} = \frac{3}{2 \tilde{M}}(d_{\mathrm{init}}^{2}-1)
\end{equation}
and therefore, from (\ref{IW820}), 
\begin{equation}
\label{IW849}
t_{\mathrm{J,SR}} - t_{\mathrm{J,init}} = \frac{3}{\tilde{M}} (d_{\mathrm{init}}-1).
\end{equation}
These analytic solutions can only be used as estimators as, strickly speaking, the slow roll approximation is 
no longer valid when $\phi \approx 0$.

Finally, we check on the small time limit of $H_{\mathrm{J}}(t_{\mathrm{J}})$. Defining
$\Delta t_{\mathrm{J}}\equiv t_{\mathrm{J}}-t_{\mathrm{J,init}}$, then
\begin{equation}
\label{IW855d}
\tau(t_{\mathrm{J}})=d_{\mathrm{init}} -\frac{\tilde{M}}{3} \Delta t_{\mathrm{J}}
\end{equation}
and, from (\ref{IW825d}),
\begin{eqnarray}
\label{IW855e}
H_{\mathrm{J}}(t_{\mathrm{J}}) & = & \frac{\tilde{M}}{2}\tau(t_{\mathrm{J}})\left[1-\frac{1}{3}\tau(t_{\mathrm{J}})^{-2}\right]
\nonumber  \\
& \approx &  H_{\mathrm{J,init}} -\frac{\tilde{M}^{2}}{6}\Delta t_{\mathrm{J}} + O(\Delta t_{\mathrm{J}}^{2})  ,
\end{eqnarray}
which agrees with the slow-roll result (\ref{IW107b}).

The calculation of gravitational baryogenesis requires $d R_{\mathrm{J}}(t_{\mathrm{J}})/d t_{\mathrm{J}}$. 
From (\ref{IW818}),
\begin{eqnarray}
\label{IW552g1}
\frac{dR_{\mathrm{J}}(t_{\mathrm{J}})}{d t_{\mathrm{J}}} & = & 3 \tilde{M}^{2}\frac{d [\tau^{2}(t_{\mathrm{J}})-1]]}{d t_{\mathrm{J}}}
\nonumber  \\
& = & -2 \tilde{M}^{3} \tau(t_{\mathrm{J}}).
\end{eqnarray}
Also required is an estimate for the decoupling temperature $T_{\mathrm{D}}$. Previous studies~\cite{ABS2010,MN2012} assume
the gravitationally produced particles totally convert into radiation in thermal equilibrium with a temperature $T$,
that is, $\rho_{\mathrm{J}}(t_{\mathrm{J}})=\rho_{\mathrm{rad}}(t_{\mathrm{J}})$, where $T$ is given by (\ref{IWg104a}):
\begin{eqnarray}
\label{IW552h}
T & = & \left[\frac{30 \rho_{\mathrm{rad}}(t_{\mathrm{J}})}{g_{\ast}(T) \pi^{2}}\right]^{1/4}
\nonumber  \\
& = & \tilde{M} (1-6 \xi)^{1/2}\left[\frac{45}{64 \pi^{3}}\right]^{1/4}\left[\frac{-e^{3\tau^{2}}I_{\tau}}{\tau^{2}}\right]^{1/4}.
\end{eqnarray}
Under this assumption (\ref{IW552h}) can be used for $T_{\mathrm{D}}$.

We will also make this thermalization assumption in order to investigate the possibility of gravitational baryogenesis in the 
$f(R)$ models considered. However, some discussion of the validity of this assumption during the inflationary era is warranted. 

The conversion of the gravitationally produced particles into a relativistic non-Abelian plasma of SM particles is usually 
modelled as occurring via the production of a new, superheavy and unstable particle, but direct conversion via the Higgs 
particle has also been studied~\cite{FB2017}. Rapid thermalization of this plasma requires a time~\cite{KM2011,FB2017} 
$t_{\mathrm{Eq}} \sim 1/(\alpha^{2} T_{\mathrm{Eq}})$ where $T_{\mathrm{Eq}}$ is the temperature of the system when thermal
equilibrium is established and $\alpha =g^{2}/(4\pi)$ with $g^{2} \simeq 0.3$ for a weakly coupled plasma of SM particles.
Hence we require the thermalization rate to satisfy $\Gamma_{\mathrm{th}} = \alpha^{2} T > H_{\mathrm{J}}$.

Gravitational baryogenesis also requires the baryon-violating processes to be in thermal equilibrium, that is, the interaction 
rate $\Gamma_{\mathrm{B}}$ of the baryon-violating processes must be greater than the expansion rate $H_{\mathrm{J}}$
until decoupling at the temperature $T_{\mathrm{D}}$.

If these conditions $\Gamma_{\mathrm{th}} > H_{\mathrm{J}}$ and $\Gamma_{\mathrm{B}} > H_{\mathrm{J}}$ are satisfied then the
Baryon Asymmetry Factor during this slow roll era is
\begin{equation}
\label{IW552j}
\eta = -\left[\frac{15 g_{B}}{4 \pi^{2} g_{\ast}(T)}\right] \frac{dR_{\mathrm{J}}/dt_{\mathrm{J}}}{M_{\ast}^{2}k_{\mathrm{B}}T}
\end{equation}
which, using (\ref{IW552g1}) and (\ref{IW552h}), gives (with $k_{\mathrm{B}}=1$)
\begin{eqnarray}
\label{IW552k}
\eta & = & \left[\frac{30 g_{B}}{4 \pi^{2} g_{\ast}(T)}\right] \frac{\tilde{M}^{3}}{M_{\ast}^{2}}
\frac{\tau(t_{\mathrm{J}})}{T}
\nonumber  \\
& = & \left[\frac{30 g_{B}}{2 \pi^{2} g_{\ast}(T)}\right]  \left(\frac{\tilde{M}}{M_{\ast}}\right)^{2} 
\left[\frac{64 \pi^{3}e^{-3\tau^{2}}}{45(-I_{\tau})}\right]^{1/4} \tau^{3/2}(t_{\mathrm{J}}).
\end{eqnarray}

The calculations are undertaken using the dimensionless units~\cite{MN2012} 
\begin{eqnarray}
\label{IW922}
\tilde{t} & = & \tilde{M}t, \,\,\tilde{\phi}=\kappa \phi, \,\, \tilde{U} = \kappa^{2}U/\tilde{M}^{2},\,\,\tilde{H}=H/\tilde{M},
\nonumber  \\
\tilde{R} & = & R/\tilde{M}^{2}, \quad \tilde{\rho}_{\mathrm{rad}} =\kappa^{2}\rho_{\mathrm{rad}}/\tilde{M}^{2}.
\end{eqnarray}
The results for the slow roll variables $\tilde{\phi}$, $\tilde{R}_{\mathrm{J}}$, $\tilde{H}_{\mathrm{J}}$, and
$\frac{d\tilde{R}_{\mathrm{J}}}{d\tilde{t}_{\mathrm{J}}}$, and the quantities $\tilde{\rho}_{\mathrm{rad}}$, 
$T$ (GeV), and $ \eta$, over the slow-roll region, for equal increments in $\log_{10}\tilde{t}_{\mathrm{J}}$,
are shown in Table I. 
We take $\tilde{M}/M_{\mathrm{Pl}}=1.2 \times 10^{-5}$, assume $M_{\ast}=M_{\mathrm{Pl}}$, and that 
the initial condition $d_{\mathrm{init}} = 10.0$. This gives $\tilde{\phi}_{\mathrm{init}}=\sqrt{6}\ln d_{\mathrm{init}} =5.6$, 
reflecting the high field value required for the Starobinsky model to be consistent with CMB observations. 
From (\ref{IW825d}) this corresponds to
\begin{equation}
\label{IW552f}
H_{\mathrm{J,init}}  =  d_{\mathrm{init}}\frac{\tilde{M}}{2}[1 - \frac{1}{3 d_{\mathrm{init}}^{2}}] =4.983 \tilde{M}.
\end{equation}

The end of the slow-roll regime is, from (\ref{IW849}),
\begin{equation}
\label{IW1045}
\tilde{t}_{\mathrm{J,SR}} -\tilde{t}_{\mathrm{J,init}} =3(d_{\mathrm{init}}-1) =27.
\end{equation}
This corresponds to $t_{\mathrm{J,SR}}=\tilde{t}_{\mathrm{J,SR}}/\tilde{M}=9.240 \times 10^{-14}$ s.

\begin{table}[htbp]
\centering
\begin{tabular}{llllllll}
$\tilde{t}_{\mathrm{J}}$ & $\tilde{\phi}$ & $\tilde{R}_{\mathrm{J}}$ & $\tilde{H}_{\mathrm{J}}$ & 
$\frac{d\tilde{R}_{\mathrm{J}}}{d\tilde{t}_{\mathrm{J}}}$ & $\tilde{\rho}_{\mathrm{rad}}$ & $T$ (GeV) & $ \eta$ \\
\hline
1.000 & 5.557 & 277.3 & 4.816 & -19.33 & 1.705(-8) & 3.959(13)  & 1.463(-11) \\
1.585 & 5.507 & 266.1 & 4.718 & -18.94 & 1.603(-8) & 3.900(13) & 1.455(-11) \\
1.995 & 5.472 & 258.4 & 4.650 & -18.67 & 1.534(-8) & 3.856(13) & 1.453(-11)  \\
2.512 & 5.426 & 248.9 & 4.563 & -18.33 & 1.450(-8) & 3.802(13) & 1.444(-11) \\
3.163 & 5.367 & 237.1 & 4.454 & -17.89 & 1.350(-8) & 3.734(13) & 1.435(-11) \\
3.981 & 5.291 & 222.7 & 4.317 & -17.35 & 1.230(-8) & 3.647(13) & 1.424(-11)  \\
5.012 & 5.192 & 205.1 & 4.145 & -16.66 & 1.086(-8) & 3.537(13) & 1.411(-11)  \\
6.310 & 5.062 & 184.1 & 3.927 & -15.79 & 9.237(-9) & 3.397(13) & 1.393(-11) \\
7.943 & 4.887 & 159.2 & 3.654 & -14.70 & 7.433(-9) & 3.217(13) & 1.369(-11)  \\
10.00 & 4.647 & 130.3 & 3.308 & -13.33 & 5.516(-9) & 2.986(13) & 1.337(-11)  \\
12.59 & 4.307 & 98.05 & 2.873 & -11.61 & 3.608(-9) & 2.685(13) & 1.295(-11)  \\
15.85 & 3.800 & 63.75 & 2.323 & -9.434 & 1.903(-9) & 2.288(13) & 1.235(-11)  \\
19.95 & 2.961 & 30.65 & 1.625 & -6.698 & 6.457(-10) & 1.747(13) & 1.149(-11) \\
25.12 & 1.192 & 4.942 & 0.7111 & -3.254 & 5.123(-11) & 9.270(12) & 1.051(-11) \\
\end{tabular}
\caption{\label{tab:anals}Results for the slow-roll era of the Starobinsky model using analytic approximations,
assuming $d_{\mathrm{init}}=10$.}
\end{table}

The calculated values $\eta  \sim (1.051 - 1.4630) \times 10^{-11}$ are quite close to the observed value 
$\eta_{\mathrm{obs}} = 8.65 \times 10^{-11}$. Since $\eta \propto (\tilde{M}/M_{\ast})^{2}$, reducing $M_{\ast}$ 
slightly from $M_{\mathrm{Pl}}$ to $0.4 M_{\mathrm{Pl}}$ would bring the calculated values into agreement with the 
observed value. 

However, if the asymmetry is generated before the end of inflation, then the baryon density is diluted as 
$n_{\mathrm{B}} \propto a^{-3}$ and  reheating subsequently injects entropy\footnote{The author wishes to thank the anonymous
referee for bringing this point to his attention}. This slow roll value of $n_{\mathrm{B}}/s$ cannot, therefore, be compared
directly with the observed late-time baryon asymmetry, and should be diluted by the factor
\begin{equation}
\label{IWe1}
\Delta = \frac{s(t_{\mathrm{J,D}}) a^{3}_{\mathrm{J}}(t_{\mathrm{J,D}})}{s(t_{\mathrm{J,end}}) a^{3}_{\mathrm{J}}(t_{\mathrm{J,end}})}
= \frac{g_{\ast}(T_{\mathrm{D}}) T^{3}_{\mathrm{D}}}{g_{\ast}(T_{\mathrm{end}}) T^{3}_{\mathrm{end}}}
\frac{n_{\mathrm{B}}(t_{\mathrm{J,end}})}{n_{\mathrm{B}}(t_{\mathrm{J,D}})},
\end{equation}
where, for production $N$ e-folds before the end of inflation,  
$n_{\mathrm{B}}(t_{\mathrm{J,end}}) =n_{\mathrm{B}}(t_{\mathrm{J,D}}) e^{-3N}$. 
Assuming inflation ends at the end of the slow roll era, then $T^{3}_{\mathrm{D}}/T^{3}_{\mathrm{end}} \lesssim 10$, 
and the dilution factor is $\Delta \lesssim 10 e^{-3N}$. The results in Table 1 therefore only apply if baryogenesis occurs 
at the end of the inflation era.

\subsection{New power-law model}

Odintsov and Oikonomou~\cite{OO2025} have proposed a new power-law model
\begin{equation}
\label{IW961a0}
f(R_{\mathrm{J}}) = \frac{16 c_{1}}{(4+k)(8+k)} R_{\mathrm{J}}^{2+k/4} + c_{2}R + c_{3}
\end{equation}
where $c_{i}$ are dimensionfull constants and $k=[-0.038,-0.022]$ to fit Planck and ACT data.

In order to develop analytic approximations of general applicability, we consider the more general power-law model
\begin{equation}
\label{IW961a1}
f(R_{\mathrm{J}}) = \lambda R_{\mathrm{J}}^{2+\epsilon} + c_{2}R + c_{3}.
\end{equation}

For this model
\begin{eqnarray}
\label{IW961b}
\Omega^{2}&  = & \frac{df(R_{\mathrm{J}})}{d R_{\mathrm{J}}}
\nonumber  \\
& = & \lambda (2+\epsilon) R_{\mathrm{J}}^{1+\epsilon} + c_{2}
\nonumber \\
& = & e^{\sqrt{2/3}\kappa \phi}.
\end{eqnarray}

The scalaron potential is
\begin{equation}
\label{IW961c}
U(\phi) = \frac{\Omega^{2} R_{\mathrm{J}}(\Omega) -f[R_{\mathrm{J}}(\Omega)]}{2 \kappa^{2} \Omega^{4}}
\end{equation}
where, from (\ref{IW961b}),
\begin{equation}
\label{IW961e}
R_{\mathrm{J}}(\Omega) = \left[\frac{1}{\lambda (2+\epsilon)}(\Omega^{2}-c_{2})\right]^{1/(1+\epsilon)}.
\end{equation}
This gives
\begin{equation}
\label{IW961h2}
U(\phi) = \beta \Omega^{-2\alpha} (1-c_{2}\Omega^{-2})^{2-\alpha}- \frac{c_{3}}{2 \kappa^{2} \Omega^{4}}
\end{equation}
where
\begin{equation}
\label{IW961h1}
\alpha \equiv \frac{\epsilon}{1+\epsilon},
\end{equation}
and
\begin{equation}
\label{IW961h}
\beta \equiv \frac{1}{2 \kappa^{2}}\left[\frac{1}{\lambda(2+\epsilon)}\right]^{1/(1+\epsilon)}
\left(\frac{1+\epsilon}{2+\epsilon}\right).
\end{equation}
Note that (\ref{IW961e}) requires $\Omega^{2}\geq c_{2}$, that is 
\begin{equation}
\label{IW961e1}
\phi \geq \frac{1}{\kappa} \sqrt{\frac{3}{2}} \ln c_{2}.
\end{equation}

Note that, for the Starobinsky model, $\epsilon=0$, $\lambda=1/6\tilde{M}^{2}$, $c_{2}=1$ and $c_{3}=0$, and (\ref{IW961h2}) reduces to
(\ref{IW199}).

This gives
\begin{eqnarray}
\label{IW961i}
\frac{dU}{d \Omega} & = & 2 \beta \Omega^{-\alpha}(1-c_{2}\Omega^{-2})^{(2-\alpha)/2}
\nonumber  \\
& & \times [-\alpha \Omega^{-\alpha-1}(1-c_{2}\Omega^{-2})^{1-\alpha/2}
+ (2-\alpha)\Omega^{-\alpha}(1-c_{2}\Omega^{-2})^{-\alpha/2} c_{2}\Omega^{-3}]
\nonumber  \\
& & +\frac{2c_{3}}{\kappa^{2}\Omega^{5}}.
\end{eqnarray}
If we choose $c_{3}=0$, then
\begin{equation}
\label{IW961i1}
\frac{dU}{d \Omega} = 2 \sqrt{\beta} \sqrt{U(\Omega)} \Omega^{-1}(\Omega^{2}-c_{2})^{-\alpha/2}[2 c_{2}\Omega^{-2} - \alpha]
\end{equation}
and therefore
\begin{equation}
\label{IW961k}
\frac{dU}{d \phi}  =  2 \kappa \sqrt{\frac{\beta}{6}} (2c_{2}\Omega^{-2}- \alpha)(\Omega^{2}-c_{2})^{-\alpha/2}
\sqrt{U(\phi)}.
\end{equation}

The slow-roll equation (\ref{IW802}) is then
\begin{eqnarray}
\label{IW961n}
\frac{d \phi}{d t_{\mathrm{E}}} & = & - \frac{dU/d\phi}{\kappa \sqrt{3U(\phi)}}
\nonumber \\
& = & -\sqrt{\frac{2 \beta}{9}}(2c_{2}\Omega^{-2} - \alpha)(\Omega^{2}-c_{2})^{-\alpha/2}.
\end{eqnarray}

An approximate solution can be obtained to (\ref{IW961n}) by assuming $\Omega^{2} \gg c_{2}$, and $c_{2}\Omega^{-2} \gg |\alpha|$.
For this case, (\ref{IW961n}) becomes 
\begin{equation}
\label{IW1000}
\frac{d \phi}{d t_{\mathrm{E}}} \approx  - 2 c_{2}\sqrt{\frac{2 \beta}{9}} \Omega^{-(\alpha +2)}. 
\end{equation}
Integrating
\begin{equation}
\label{IW1001}
\frac{d t_{\mathrm{E}}}{d \phi} = - \frac{1}{2c_{2}} \sqrt{\frac{9}{2 \beta}} e^{(\alpha+2)\kappa \phi /\sqrt{6}}
\end{equation}
gives
\begin{equation}
\label{IW1002}
t_{\mathrm{E}} - t_{\mathrm{E,init}} = -\gamma^{-1} \left[e^{(\alpha+2)\kappa \phi /\sqrt{6}} - \hat{d}^{2}_{\mathrm{init}}\right],
\end{equation}
where
\begin{equation}
\label{IW1003}
\gamma \equiv  \frac{2}{3}  \kappa c_{2} (\alpha+2) \sqrt{\frac{\beta}{3}}
\end{equation}
and
\begin{equation}
\label{IW1004}
\hat{d}^{2}_{\mathrm{init}} \equiv e^{(\alpha+2)\kappa \phi_{\mathrm{init}}/\sqrt{6}}.
\end{equation}
Thus we obtain
\begin{equation}
\label{IW1005}
\phi(t_{\mathrm{E}}) = \frac{2}{(\alpha +2) \kappa}\sqrt{\frac{3}{2}} \ln \left[\hat{d}^{2}_{\mathrm{init}}
- \gamma (t_{\mathrm{E}} - t_{\mathrm{E,init}})\right].
\end{equation}

This immediately gives
\begin{equation}
\label{IW1006}
\frac{d \phi}{d t_{\mathrm{E}}} = -\gamma \frac{2}{(\alpha +2) \kappa} \sqrt{\frac{3}{2}}
\left[\hat{d}^{2}_{\mathrm{init}} -\gamma (t_{\mathrm{E}} - t_{\mathrm{E,init}})\right]^{-1}
\end{equation}
and therefore the slow-roll equation (\ref{IW801})
\begin{eqnarray}
\label{IW1007}  
\frac{d H_{\mathrm{E}}}{d t_{\mathrm{E}}} & = & -\frac{\kappa^{2}}{2} \left(\frac{d \phi}{d t_{\mathrm{E}}}\right)^{2}
\nonumber  \\
& = & - 3 \left[\frac{\gamma}{(\alpha+2)}\right]^{2}
\left[\hat{d}^{2}_{\mathrm{init}} -\gamma (t_{\mathrm{E}} - t_{\mathrm{E,init}})\right]^{-2}.
\nonumber  \\
\end{eqnarray}
Defining 
\begin{equation}
\label{IW1008}
\hat{T} \equiv \hat{d}^{2}_{\mathrm{init}} -\gamma (t_{\mathrm{E}} - t_{\mathrm{E,init}})
\end{equation}
then $d\hat{T}/dt_{\mathrm{E}} =-\gamma $ and (\ref{IW1007}) becomes
\begin{equation}
\label{IW1009}
\frac{d H_{\mathrm{E}}}{d \hat{T}} = \frac{3\gamma}{(\alpha+2)^{2}} \tilde{T}^{-2}.
\end{equation}
Integrating, and noting that $\tilde{T}_{\mathrm{init}}=\tilde{d}^{2}_{\mathrm{init}}$, gives
\begin{equation}
\label{IW1010}
H_{\mathrm{E}} = H_{\mathrm{E,init}} 
-\frac{3\gamma}{(\alpha+2)^{2}}\left[\frac{1}{\hat{d}^{2}_{\mathrm{init}}} - 
\frac{1}{\hat{d}^{2}_{\mathrm{init}} -\gamma (t_{\mathrm{E}} - t_{\mathrm{E,init}})}\right].
\end{equation}
Setting
\begin{equation}
\label{IW1011}
H_{\mathrm{E,init}} = \frac{3\gamma}{(\alpha+2)^{2}}\left[1-\frac{1}{\hat{d}^{2}_{\mathrm{init}}}\right]
\end{equation}
then
\begin{eqnarray}
\label{IW1012}
H_{\mathrm{E}}(t_{\mathrm{E}}) & = & \xi \left[1 - \frac{1}{\hat{d}^{2}_{\mathrm{init}} 
-\gamma (t_{\mathrm{E}} - t_{\mathrm{E,init}})}\right].
\end{eqnarray}
where
\begin{equation}
\label{IW1012a}
\xi \equiv \frac{3\gamma}{(\alpha+2)^{2}}.
\end{equation}

This result (\ref{IW1012}) for $H_{\mathrm{E}}(t_{\mathrm{E}})$ has the same form as that for the Starobinsky model with
$\xi \rightarrow \tilde{M}/2$, giving $H_{\mathrm{E}} > 0$ and decreasing with $t_{\mathrm{E}}$ as desired.

Integrating  $H_{\mathrm{E}}(t_{\mathrm{E}}) =d(\ln a_{\mathrm{E}})/d t_{\mathrm{E}}$ gives
\begin{equation}
\label{IW1014}
a_{\mathrm{E}}(t_{\mathrm{E}})= a_{\mathrm{a,init}} e^{\xi(t_{\mathrm{E}} - t_{\mathrm{E,init}})}
\left[1-\frac{\gamma}{\hat{d}^{2}_{\mathrm{init}}}(t_{\mathrm{E}} - t_{\mathrm{E,init}})\right]^{\xi/\gamma}.
\end{equation}

The transformation between $t_{\mathrm{E}}$ and $t_{\mathrm{J}}$ can be obtained from
\begin{eqnarray}
\label{IW1017}
\frac{d t_{\mathrm{J}}}{dt _{\mathrm{E}}} & = & \Omega^{-1}(t_{\mathrm{E}})
\nonumber  \\
& = & e^{-\kappa \phi(t_{\mathrm{E}})/\sqrt{6}}
\nonumber  \\
& = & [\hat{d}^{2}_{\mathrm{init}}-\gamma (t_{\mathrm{E}} - t_{\mathrm{E,init}})]^{-1/(\alpha +2)}.
\end{eqnarray}
Integrating gives
\begin{equation}
\label{IW1019}
t_{\mathrm{J}}-t_{\mathrm{J,init}} =  \frac{1}{\hat{\alpha}\gamma}\left\{d_{\mathrm{init}}^{2\hat{\alpha}} 
-[\hat{d}^{2}_{\mathrm{init}}-\gamma (t_{\mathrm{E}} - t_{\mathrm{E,init}})]^{\hat{\alpha}} \right\}
\end{equation}
where
\begin{equation}
\label{IW1020}
\hat{\alpha} \equiv \frac{(\alpha+1)}{(\alpha +2)}.
\end{equation}
Inverting this gives
\begin{equation}
\label{IW1021}
\hat{d}^{2}_{\mathrm{init}}-\gamma (t_{\mathrm{E}} - t_{\mathrm{E,init}})]^{\hat{\alpha}}
=  \hat{d}_{\mathrm{init}}^{2\hat{\alpha}} - \gamma \hat{\alpha}(t_{\mathrm{J}}-t_{\mathrm{J,init}}),
\end{equation}
and therefore 
\begin{equation}
\label{IW1022}
t_{\mathrm{E}} - t_{\mathrm{E,init}} = \frac{1}{\gamma}\left\{\hat{d}_{\mathrm{init}}^{2} 
-[\hat{d}_{\mathrm{init}}^{2\hat{\alpha}} -
\gamma \hat{\alpha}(t_{\mathrm{J}}-t_{\mathrm{J,init}})]^{1/\hat{\alpha}}.
\right\}.
\end{equation}

Defining
\begin{equation}
\label{IW1023a} 
\hat{\tau}(t_{\mathrm{E}}) \equiv [\hat{d}^{2}_{\mathrm{init}}-\gamma (t_{\mathrm{E}} - t_{\mathrm{E,init}})]^{1/(\alpha +2)}
\end{equation}
and
\begin{equation}
\label{IW1023b}
\hat{\tau}(t_{\mathrm{J}}) \equiv [\hat{d}_{\mathrm{init}}^{2\hat{\alpha}} - \gamma \hat{\alpha}
(t_{\mathrm{J}}-t_{\mathrm{J,init}})]^{1/(\alpha+1)}
\end{equation}
then, from (\ref{IW1017}) and (\ref{IW1021}),
\begin{equation}
\label{IW1023}
\Omega (t_{\mathrm{E}}) = \hat{\tau}(t_{\mathrm{E}}) =  \hat{\tau}(t_{\mathrm{J}}) =  \Omega (t_{\mathrm{J}}).
\end{equation}
Hence
\begin{eqnarray}
\label{IW1024}
R_{\mathrm{J}}(t_{\mathrm{E}}) & = &\left[\frac{1}{\lambda(2+\epsilon)}(\Omega^{2}(t_{\mathrm{E}})-c_{2})\right]^{1/(1+\epsilon)}
\nonumber  \\
& = & \left[\frac{1}{\lambda(2+\epsilon)} (\hat{\tau}^{2}(t_{\mathrm{E}}) -c_{2})\right]^{1/(1+\epsilon)},
\end{eqnarray}
and
\begin{equation}
\label{IW1024a}
R_{\mathrm{J}}(t_{\mathrm{J}})  =  \left[\frac{1}{\lambda(2+\epsilon)} (\hat{\tau}^{2}(t_{\mathrm{J}}) -c_{2})\right]^{1/(1+\epsilon)}.
\end{equation}

Also, (\ref{IW1005}) becomes
\begin{equation}
\label{IW1023a1}
\phi(t_{\mathrm{E}})=\frac{\sqrt{6}}{\kappa} \ln [\hat{\tau}(t_{\mathrm{E}})]
\end{equation}
and (\ref{IW1012}) becomes
\begin{equation}
\label{IW1023a2}
H_{\mathrm{E}}(t_{\mathrm{E}}) = \xi \left\{1-[\hat{\tau}(t_{\mathrm{E}})]^{-(\alpha+2)}\right\}.
\end{equation}
Noting that
\begin{equation}
\label{IW1023a3}
\frac{d \phi (t_{\mathrm{E}})}{d t_{\mathrm{E}})} = -\frac{\gamma}{(\alpha+2)}\frac{\sqrt{6}}{\kappa} 
[\hat{\tau}(t_{\mathrm{E}})]^{-(\alpha+2)},
\end{equation}
then
\begin{equation}
\label{IW1023a4}
H_{\mathrm{J}}(t_{\mathrm{E}}) = \xi  \hat{\tau}(t_{\mathrm{E}})
\left\{1-\left[1-\frac{\gamma}{(\alpha+2)\xi}\right]\hat{\tau}(t_{\mathrm{E}})]^{-(\alpha+2)}\right\}
\end{equation}
and therefore
\begin{equation}
\label{IW1023a5}
H_{\mathrm{J}}(t_{\mathrm{J}}) = \xi  \hat{\tau}(t_{\mathrm{J}})
\left\{1-\left[1-\frac{\gamma}{(\alpha+2)\xi}\right]\hat{\tau}(t_{\mathrm{J}})]^{-(\alpha+2)}\right\}.
\end{equation}
Since $\gamma/(\alpha+2)\xi=(\alpha+2)/3$,  (\ref{IW1023a5}) reduces to (\ref{IW825d}) for $\alpha=0$.

The expression (\ref{IW1014}) for $a_{\mathrm{E}}(t_{\mathrm{E}})$ can be rewritten using (\ref{IW1023a}).
Noting that $\xi/\gamma = 3/(\alpha+2)^{2}$, then
\begin{equation}
\label{IW1024a0}
[\hat{d}^{2}_{\mathrm{init}}-\gamma (t_{\mathrm{E}} - t_{\mathrm{E,init}})]^{\xi/\gamma}
= \hat{\tau}(t_{\mathrm{E}})^{3/(\alpha+2)}.
\end{equation}
Also, from (\ref{IW1002}),
\begin{equation}
\label{IW1024b}
\xi (t_{\mathrm{E}} - t_{\mathrm{E,init}}) = \frac{3}{(\alpha+2)^{2}}
\left[\hat{d}^{2}_{\mathrm{init}}\right] -\hat{\tau}(t_{\mathrm{E}})^{\alpha+2},
\end{equation}
so that
\begin{equation}
\label{IW1024c}
e^{\xi (t_{\mathrm{E}} - t_{\mathrm{E,init}})} = e^{\frac{3}{(\alpha+2)^{2}}\hat{d}^{2}_{\mathrm{init}}}
e^{-\frac{3}{(\alpha+2)^{2}}\hat{\tau}(t_{\mathrm{E}})^{\alpha+2}}.
\end{equation}
Hence we obtain
\begin{equation}
\label{IW1024d}
\frac{a_{\mathrm{E}}(t_{\mathrm{E}})}{a_{\mathrm{E}}(t_{\mathrm{E,init}})} =  e^{\frac{3}{(\alpha+2)^{2}}\hat{d}^{2}_{\mathrm{init}}}
e^{-\frac{3}{(\alpha+2)^{2}}\hat{\tau}(t_{\mathrm{E}})^{\alpha+2}}
\left[\frac{\hat{\tau}(t_{\mathrm{E}})^{\alpha+2}}{\hat{d}^{2}_{\mathrm{init}}}\right]^{3/(\alpha+2)^{2}}.
\end{equation}

Using
\begin{equation}
\label{IW1023a0}
\frac{d \hat{\tau}(t_{\mathrm{J}})}{d t_{\mathrm{J}}} =  -\frac{\gamma}{\alpha+2} \hat{\tau}^{-\alpha}(t_{\mathrm{J}}),
\end{equation}
then
\begin{equation}
\label{IW1025}
\frac{d R_{\mathrm{J}}(t_{\mathrm{J}})}{d t_{\mathrm{J}}} =  - \zeta [\hat{\tau}^{2}(t_{\mathrm{J}}) -c_{2}]^{-\alpha}
\hat{\tau}^{1-\alpha}(t_{\mathrm{J}}),
\end{equation}
where the prefactor is
\begin{equation}
\label{IW1025a} 
\zeta \equiv \frac{2\gamma}{\alpha+2} \left[\frac{1}{\lambda(2+\epsilon)}\right]^{-\alpha} 
\frac{1}{\lambda(1+\epsilon)(2+\epsilon)}. 
\end{equation}

The gravitationally produced particle density is, from (\ref{IW823}) and assuming $\xi =0$,
\begin{eqnarray}
\label{IW1026}
\rho_{\mathrm{J}}(t_{\mathrm{J}})&  = & \frac{g_{\ast}(T)\tilde{M}}{1152 \pi a_{\mathrm{J}}(t_{\mathrm{J}})^{4}}  
\int^{t_{\mathrm{J}}}_{-\infty} 
a_{\mathrm{J}}^{4}(t^{\prime}_{\mathrm{J}}) R_{\mathrm{J}}^{2}(t^{\prime}_{\mathrm{J}}) dt^{\prime}_{\mathrm{J}}
\nonumber  \\
& = & \frac{g_{\ast}(T)\tilde{M}}{1152 \pi} 
\left[\frac{\Omega (t_{\mathrm{J}})}{a_{\mathrm{E}}(t_{\mathrm{J}})}\right]^{4}
\nonumber  \\
& & \times \int^{t_{\mathrm{J}}}_{-\infty} 
\left[\Omega^{-1} (t^{\prime}_{\mathrm{E}})a_{\mathrm{E}}(t^{\prime}_{\mathrm{E}})\right]^{4}
R_{\mathrm{J}}^{2}(t^{\prime}_{\mathrm{J}})
\Omega^{-1} (t^{\prime}_{\mathrm{E}}) dt^{\prime}_{\mathrm{E}}.
\end{eqnarray}

In the slow-roll approximation, we can use (\ref{IW1024d}) for $a_{\mathrm{E}}(t_{\mathrm{E}})$, and (\ref{IW1024})
for $R_{\mathrm{J}}(t_{\mathrm{E}})$, to obtain
\begin{eqnarray}
\label{IW1027}
\rho_{\mathrm{J}}(t_{\mathrm{J}})&  = & \frac{g_{\ast}(T)\tilde{M}}{1152 \pi} \left(\frac{1}{\lambda(2+\epsilon)}\right)^{2/(1+\epsilon)}
\left[\frac{a_{\mathrm{E}}(t_{\mathrm{E,init}})}{a_{\mathrm{E}}(t_{\mathrm{J}})}\right]^{4} \hat{\tau}^{4}(t_{\mathrm{J}})
\nonumber  \\
& & \times e^{\frac{12}{(\alpha+2)^{2}}\hat{d}^{2}_{\mathrm{init}}}
\left[\hat{d}^{2}_{\mathrm{init}}\right]^{-12/(\alpha+2)^{2}}
\nonumber  \\
& & \times \int^{t_{\mathrm{J}}}_{t_{\mathrm{J,init}}} 
e^{-\frac{12}{(\alpha+2)^{2}}\hat{\tau}(t^{\prime}_{\mathrm{E}})^{\alpha+2}}
\hat{\tau}^{-5}(t^{\prime}_{\mathrm{E}}) 
\hat{\tau}(t^{\prime}_{\mathrm{E}})^{\frac{12}{(\alpha+2)}}
[\hat{\tau}^{2}(t^{\prime}_{\mathrm{E}}) - c_{2}]^{2/(1+\epsilon)} dt^{\prime}_{\mathrm{E}}.
\end{eqnarray}
Changing the integration variable to $\hat{\tau}(t^{\prime}_{\mathrm{E}})$ using 
\begin{equation}
\label{IW1028}
\frac{d t^{\prime}_{\mathrm{E}}}{d \hat{\tau}} = -\left(\frac{\alpha+2}{\gamma}\right) \hat{\tau}^{\alpha +1}
\end{equation}
and, from (\ref{IW1024d}),
\begin{equation}
\label{IW1029}
\left[\frac{a_{\mathrm{E}}(t_{\mathrm{J,init}})}{a_{\mathrm{E}}(t_{\mathrm{J}})}\right]^{4} = 
e^{-\frac{12}{(\alpha+2)^{2}}\hat{d}^{2}_{\mathrm{init}}} e^{\frac{12}{(\alpha+2)^{2}}\hat{\tau}(t_{\mathrm{J}})^{\alpha+2}}
\left[\hat{d}^{2}_{\mathrm{init}}\right]^{\frac{12}{(\alpha+2)^{2}}} \,\hat{\tau}(t_{\mathrm{J}})^{-\frac{12}{(\alpha+2)}},
\end{equation}
(since $t_{\mathrm{J,init}}=t_{\mathrm{E,init}}$) then the energy density is
\begin{eqnarray}
\label{IW1030}
\rho_{\mathrm{J}}(t_{\mathrm{J}})&  = & -\frac{g_{\ast}(T)\tilde{M}}{1152 \pi} 
\left(\frac{1}{\lambda(2+\epsilon)}\right)^{2/(1+\epsilon)}\hat{\tau}^{4}(t_{\mathrm{J}})
e^{\frac{12}{(\alpha+2)^{2}}\hat{\tau}(t_{\mathrm{J}})^{\alpha+2}}
\hat{\tau}(t_{\mathrm{J}})^{\frac{-12}{(\alpha+2)}} \left(\frac{\alpha+2}{\gamma}\right)
\nonumber  \\
& & \times  \int^{\hat{\tau}(t_{\mathrm{J}})}_{\hat{\tau}(t_{\mathrm{J,init}})}
e^{-\frac{12}{(\alpha+2)^{2}}\hat{\tau}(t^{\prime}_{\mathrm{E}})^{\alpha+2}}
\hat{\tau}(t^{\prime}_{\mathrm{E}})^{\frac{(\alpha^{2}-2 \alpha+4)}{(\alpha+2)}} 
[\hat{\tau}^{2}(t^{\prime}_{\mathrm{E}}) - c_{2}]^{2/(1+\epsilon)} d \hat{\tau}(t^{\prime}_{\mathrm{E}}),
\end{eqnarray}
where 
\begin{equation}
\label{IW1031a}
\hat{\tau}(t_{\mathrm{J,init}}) = \hat{d}_{\mathrm{init}}^{2/(\alpha+2)}.
\end{equation}

Defining the integral
\begin{equation}
\label{IW1031}
I_{\tau}(a,b,c,d,e) \equiv \int^{\tau}_{\tau_{\mathrm{init}}} e^{-a\tau^{b}} \tau^{c} [\tau^{2}-d]^{e} d \tau
\end{equation}
then
\begin{equation}
\label{IW1032}
\rho_{\mathrm{J}}(t_{\mathrm{J}})  =  -\frac{g_{\ast}(T)}{1152 \pi} \zeta_{\mathrm{rad}} \Theta (t_{\mathrm{J}}),
\end{equation}
where
\begin{equation}
\label{IW1032a}
\zeta_{\mathrm{rad}} \equiv \tilde{M} \left(\frac{\alpha+2}{\gamma}\right) \left(\frac{1}{\lambda(2+\epsilon)}\right)^{2/(1+\epsilon)}
\end{equation}
and
\begin{equation}
\label{IW1032b}
\Theta (t_{\mathrm{J}})  \equiv \hat{\tau}^{4}(t_{\mathrm{J}}) e^{\frac{12}{(\alpha+2)^{2}}\hat{\tau}(t_{\mathrm{J}})^{\alpha+2}}
\hat{\tau}(t_{\mathrm{J}})^{\frac{-12}{(\alpha+2)}} 
I_{\hat{\tau}(t_{\mathrm{J}})}(a,b,c,d,e).
\end{equation}
The parameters in (\ref{IW1032b}) are
\begin{equation}
\label{IW1033}
a =  \frac{12}{(\alpha+2)^{2}}, \quad b=\alpha +2, \quad  c = \frac{\alpha^{2}-2\alpha+4}{\alpha+2},
\quad d=c_{2}, \quad e=\frac{2}{1+\epsilon}.
\end{equation}

For the Starobinsky case, $\epsilon=0$, $\alpha =0$, $\lambda=1/6\tilde{M}^{2}$, $\gamma=2\tilde{M}/3$, $\hat{\tau}=\tau$, 
this integral becomes
\begin{equation}
\label{IW1034}
I_{\tau}(3,2,2,1,2)=I_{\tau},
\end{equation}
where $I_{\tau}$, defined by (\ref{IW828}), can be analytically evaluated, with the result (\ref{IW835}). 
However, an analytic evaluation of $I_{\tau}(a,b,c,d,e)$ seems unlikely and it must be obtained by numerical quadrature.

Again assuming thermalization, the temperature of the radiation is
\begin{eqnarray}
\label{IW1041}
T & = & \left[ \frac{30 \rho_{\mathrm{rad}}}{g_{\ast}(T) \pi^{2}}\right]^{1/4}
\nonumber  \\
& = & \tilde{M} \left[\frac{5}{192 \pi^{3}}\right]^{1/4} \left[ \left(\frac{\zeta_{\mathrm{rad}}}{\tilde{M}^{4}}\right)\,
(-\Theta (t_{\mathrm{J}})) \right]^{1/4},
\end{eqnarray}
and the Baryon Asymmetry Factor is
\begin{eqnarray}
\label{IW1042}
\eta & = & - \left[\frac{15 g_{B}}{4 \pi^{2} g_{\ast}(T)}\right]
\frac{d R_{\mathrm{J}}/d t_{\mathrm{J}}}{M_{\ast}^{2} T}
\nonumber  \\
& = &  \left[\frac{15 g_{B}}{4 \pi^{2} g_{\ast}(T)}\right]\left[\frac{192 \pi^{3}}{5}\right]^{1/4}
\left(\frac{\tilde{M}}{M_{\ast}}\right)^{2}
\frac{\zeta/\tilde{M}^{3}}{[(\zeta_{\mathrm{rad}}/\tilde{M}^{4})(- \Theta (t_{\mathrm{J}}))]^{1/4}}
\nonumber  \\
& & \times [\hat{\tau}^{2}(t_{\mathrm{J}})-c_{2}]^{-\alpha}\, \hat{\tau}^{1-\alpha}(t_{\mathrm{J}})
\end{eqnarray}
where
\begin{equation}
\label{IW1036a}
\hat{\tau}(\tilde{t}_{\mathrm{J}}) = [\hat{d}_{\mathrm{init}}^{2\hat{\alpha}} - 
\frac{\gamma \hat{\alpha}}{\tilde{M}}(\tilde{t}_{\mathrm{J}}-\tilde{t}_{\mathrm{J,init}})]^{1/(\alpha+1)}.
\end{equation}

For $\alpha=0$, $\zeta/\tilde{M}^{3}=2$, $\zeta_{\mathrm{rad}}/\tilde{M}^{4}=27$,
\begin{equation}
\label{IW1043}
\Theta (t_{\mathrm{J}})  =  \hat{\tau}^{-2}(t_{\mathrm{J}}) e^{3 \hat{\tau}^{2}} I_{\tau}
\end{equation}
and (\ref{IW1041}) and (\ref{IW1042}) reduce to (\ref{IW552h}) and (\ref{IW552k}) respectively.

These general expressions can now be applied to the specific model of \cite{OO2025} by choosing
\begin{equation}
\label{IW1038a}
\lambda = \frac{16 c_{1}}{(k+4)(k+8)}, \quad \epsilon = \frac{k}{4}.
\end{equation}
We then have
\begin{equation}
\label{IW1038b}
\zeta = \frac{4 c_{2}}{3\sqrt{6}}\left[\left(\frac{1}{c_{1}}\right)^{4/(k+4)}\right]^{3/2} 
\left(\frac{k+4}{k+8}\right)^{1/2} \left(\frac{k+4}{4}\right)^{(2-k)/(k+4)},
\end{equation}
and
\begin{equation}
\label{IW1038c}
\zeta_{\mathrm{rad}} = \tilde{M} \left(\frac{\alpha+2}{\gamma}\right) \left(\frac{k+4}{4c_{1}}\right)^{8/(k+4)}.
\end{equation}

For $k=-0.03$, $\alpha=-0.0076$, $\beta = 0.2472 \kappa^{-2}c_{1}^{-1.00756}$, $\gamma = 0.3813 c_{2} c_{1}^{-0.5038}$,
$\xi =0.2880 c_{1}^{-0.5038}$, $\hat{\alpha}=0.4981$, and $\gamma \hat{\alpha}=0.1899 c_{2}c_{1}^{-0.5038}$.
The prefactors are $\zeta =0.3827 c_{2} c_{1}^{-1.5113}$ and $\zeta_{\mathrm{rad}}=5.1471 c_{2}^{-1} c_{1}^{-1.5113}$.

The model contains the undetermined integration constants $c_{1}$ and $c_{2}$. To obtain values for our calculations, 
we choose the $k=0$ limit of the model to be the Starobinsky model, that is $c_{1}=1/3\tilde{M}^{2}$, $c_{2}=1$ and $c_{3}=0$.
As $1/c_{1}$ has the dimensions $(\tilde{M}^{2})^{(k+4)/4}$, we take $1/c_{1}=3 \tilde{M}^{(k+4)/2}$.

Thus, for $k=-0.03$, $1/c_{1}=3 \tilde{M}^{1.985}$ and
\begin{equation}
\label{IW1039}
\gamma =  0.6631 \tilde{M}, \quad \gamma \hat{\alpha} = 0.3303 \tilde{M}, 
\quad \zeta  = 2.0136 \tilde{M}^{3}, \quad \zeta_{\mathrm{rad}} = 27.08 \tilde{M}^{4}.
\end{equation}
The coefficients $\zeta$ and $\zeta_{\mathrm{rad}}$ are slightly enhanced compared to the Starobinsky factors $2\tilde{M}^{3}$
and $27 \tilde{M}^{4}$, respectively.

Results of calculations for $k=-0.03$ and $\hat{d}_{\mathrm{init}}=10$ are shown in Table II. 

If the end of the slow-roll regime is estimated from $\phi(t_{\mathrm{J}})=0$, then we have from (\ref{IW1036a}),
\begin{equation}
\label{IW1046}
\hat{\tau}(\tilde{t}_{\mathrm{J,SR}}) = [\hat{d}_{\mathrm{init}}^{2\hat{\alpha}} - 
\frac{\gamma \hat{\alpha}}{\tilde{M}}(\tilde{t}_{\mathrm{J,SR}}-\tilde{t}_{\mathrm{J,init}})]^{1/(\alpha+1)} =1,
\end{equation}
that is
\begin{eqnarray}
\label{IW1047}
\tilde{t}_{\mathrm{J,SR}} -\tilde{t}_{\mathrm{J,init}} & = & 
\frac{\tilde{M}}{\gamma \hat{\alpha}}(\hat{d}_{\mathrm{init}}^{2\hat{\alpha}}-1)
\nonumber  \\
& = & 3.028 (\hat{d}_{\mathrm{init}}^{0.9962}-1) = 27.
\end{eqnarray}
Thus the slow-roll regime for the Odintsov and Oikonomou model is the same as for the Starobinsky model.

The slow roll Baryon Asymmetry Factor $\eta$ varies from $1.53 \times 10^{-11}$ to 
$1.06 \times 10^{-11}$, comparable to the values for the Starobinsky model. However, it should be noted that the results
depend upon the value chosen for $c_{1}$, and a future fit of the Odintsov and Oikonomou model to data could yield
enhanced values.

\begin{table}
\centering
\begin{tabular}{llllllll}
$\tilde{t}_{\mathrm{J}}$ & $\tilde{\phi}$ & $\tilde{R}_{\mathrm{J}}$ & $\tilde{H}_{\mathrm{J}}$ & 
$\frac{d\tilde{R}_{\mathrm{J}}}{d\tilde{t}_{\mathrm{J}}}$ & $\tilde{\rho}_{\mathrm{rad}}$ & $T$ (GeV) & $ \eta$ \\
\hline
1.000 & 5.578 & 292.2 & 4.868 & -20.67 & 1.873(-8) & 4.053(13) & 1.527(-11) \\
1.585 & 5.528 & 280.2 & 4.768 & -20.24 & 1.759(-8) & 3.990(13) & 1.519(-11) \\
1.995 & 5.492 & 272.0 & 4.699 & -19.94 & 1.682(-8) & 3.946(13) & 1.513(-11)  \\
2.512 & 5.446 & 261.8 & 4.611 & -19.56 & 1.588(-8) & 3.890(13) & 1.506(-11) \\
3.163 & 5.387 & 249.2 & 4.500 & -19.08 & 1.475(-8) & 3.819(13) & 1.497(-11) \\
3.981 & 5.310 & 233.8 & 4.360 & -18.48 & 1.341(-8) & 3.729(13) & 1.485(-11)  \\
5.012 & 5.211 & 215.2 & 4.185 & -17.73 & 1.184(-8) & 3.615(13) & 1.469(-11)  \\
6.310 & 5.079 & 192.8 & 3.964 & -16.78 & 1.005(-8) & 3.469(13) & 1.449(-11) \\
7.943 & 4.903 & 166.4 & 3.685 & -15.59 & 8.058(-9) & 3.283(13) & 1.422(-11)  \\
10.00 & 4.661 & 135.8 & 3.335 & -14.09 & 5.951(-9) & 3.043(13) & 1.387(-11)  \\
12.59 & 4.319 & 101.8 & 2.893 & -12.22 & 3.866(-9) & 2.732(13) & 1.339(-11)  \\
15.85 & 3.808 & 65.76 & 2.336 & -9.869 & 2.017(-9) & 2.322(13) & 1.273(-11)  \\
19.95 & 2.964 & 31.30 & 1.630 & -6.934 & 6.725(-10) & 1.765(13) & 1.177(-11) \\
25.12 & 1.185 & 4.914 & 0.709 & -3.290 & 5.123(-10) & 9.270(12) & 1.063(-11) \\
\end{tabular}
\caption{\label{tab:power}Results for the slow-roll era of the Odintsov and Oikonomou model using analytic approximations,
for $k=-0.03$ and $\hat{d}_{\mathrm{init}}=10$.}
\end{table} 

Again, if the asymmetry is generated before the end of inflation, then the baryon density is diluted by the factor  
$\Delta \lesssim 10 e^{-3N}$. The results in Table 2 therefore only apply if baryogenesis occurs at the end of inflation.

\section{Summary and conclusions}

The feasibility of gravitational baryogenesis during the slow-roll inflation era has been investigated for two $f(R)$ 
cosmologies: the $R^{2}$ model of Starobinsky ~\cite{Star1980}, and the new power-law model of Odintsov and Oikonomou~\cite{OO2025}. 
Rather than use the Jordan frame in which these 
cosmologies have been formulated, the present investigation is undertaken in the Einstein frame, which is obtained from the Jordan
frame by a conformal transformation $\Omega$ that introduces the scalaron field $\phi =\kappa^{-1}\sqrt{6}\ln \Omega$.
The motion of the scalaron is studied for the inflationary era of slow roll using analytic approximate solutions obtained 
from its potential $U(\phi)$ and, from these solutions, analytic expressions obtained for the Ricci scalar, 
its time derivative, the Hubble parameter and the scale factor of the Universe. 
These expressions for the Starobinsky model were obtained first by \cite{MN2012}, but
the expressions for the power-law model of Odintsov and Oikonomou and its generalization are new. 

The two models chosen reflect different perspectives on studying $f(R)$ gravity. The Starobinsky model is strongly motivated by 
theoretical considerations of corrections to Einstein's General Relativity arising from quantum corrections and new physics,
whereas the power-law model of Odintsov and Oikonomou is constructed purely from the slow-roll parameters of  $f(R)$ gravity 
to fit the most recent CMD observations and lacks theoretical support.

The generation of a baryon asymmetry by the interaction (\ref{IW2}) requires the Universe to be in thermal equilibrium. The present 
calculations reported in Tables I and II have assumed the baryon-number violating interactions are more rapid than the expansion 
rate of the Universe, and that the gravitationally produced particles have been totally converted into radiation in thermal equilibrium.
The analytic approximations developed for $R_{\mathrm{J}}$ and $a_{\mathrm{J}}$ allow the evaluation of $\rho_{\mathrm{rad}}(t_{\mathrm{J}}) =\rho_{\mathrm{J}}$ and the temperature $T$. Analytic results are obtained for the Starobinsky model, whereas for the power-law model, 
the results are expressed in terms of an integral $I_{\tau}$ which required numerical evaluation.

To obtain the analytic approximations for the power-law model, it was assumed that $1 \gg c_{2} \Omega^{-2} \gg |\alpha| $ in order to 
obtain an analytic solution to the slow-roll equation (\ref{IW961n}) for $\phi(t_{\mathrm{E}})$. The validity of this 
assumption was checked by numerical solution of this equation. The differential equation was integrated over the slow-roll region using an equally spaced grid of $\log_{10} \tilde{t}_{\mathrm{E}}$. An estimate of the end of the slow-roll regime in the 
Einstein frame is, from ({\ref{IW1023}),
\begin{equation}
\label{IW1553g}
\hat{\tau}(\tilde{t}_{\mathrm{E,SR}}) = 1 = [\hat{d}^{2}_{\mathrm{init}} - 
\frac{\gamma}{\tilde{M}}(\tilde{t}_{\mathrm{E,SR}}-\tilde{t}_{\mathrm{E,init}})]^{1/(\alpha+2)},
\end{equation}
which gives
\begin{equation}
\label{IW1553h}
\tilde{t}_{\mathrm{E,SR}}-\tilde{t}_{\mathrm{E,init}}  =  \frac{\tilde{M}}{\gamma}
\left[\hat{d}^{2}_{\mathrm{init}} -1\right] = 149.3.
\end{equation}
The numerical results show that the analytical approximation for $\tilde{\phi}(\tilde{t}_{\mathrm{E}})$ is extremely good for all but
the near-end of the slow-roll region, the differences being less than 1\% for most of the range. 

The analytic approximations for the two models were obtained assuming that the energy density of the created radiation is sub-dominant
compared to the effective energy density of the $f(R)$ system. This can be validated by expressing the
modified Friedmann equation (\ref{IW103}) in the more familar form
\begin{equation}
\label{IW1553i}
H_{\mathrm{J}}^{2} = \frac{\kappa^{2}}{3}[\rho_{\mathrm{grav}} + \rho],
\end{equation}
where
\begin{equation}
\label{IW1553j}
\kappa^{2} \rho_{\mathrm{grav}} \equiv \frac{1}{2}R_{\mathrm{J}} f^{\prime}(R_{\mathrm{J}}) - f(R_{\mathrm{J}}) - 
3 H_{\mathrm{J}} \dot{R_{\mathrm{J}}} f^{\prime \prime}(R_{\mathrm{J}})
- \kappa^{2} H_{\mathrm{J}}^{2} \mathcal{F}(R_{\mathrm{J}})
\end{equation}
and
\begin{equation}
\label{IW1553k}
f^{\prime}(R_{\mathrm{J}}) = 1 + \frac{\kappa^{2}}{3}\mathcal{F}(R_{\mathrm{J}}).
\end{equation}
Using the Starobinsky model to estimate $\rho_{\mathrm{grav}}$  gives
\begin{equation}
\label{IW1553l}
\rho_{\mathrm{grav}} = \frac{3}{\kappa^{2}\tilde{M}^{2}}
(\dot{H}^{2}_{\mathrm{J}} - 6 H^{2}_{\mathrm{J}}\dot{H}_{\mathrm{J}} - 2H_{\mathrm{J}}\ddot{H}_{\mathrm{J}}).
\end{equation}
Assuming slow-roll conditions, 
\begin{equation}
\label{IW1553m}
\rho_{\mathrm{grav}} \simeq -\frac{18}{\kappa^{2}\tilde{M}^{2}} H^{2}_{\mathrm{J}}\dot{H}_{\mathrm{J}} 
\end{equation}
and therefore, using (\ref{IW825d}) for $H_{\mathrm{J}}$,
\begin{equation}
\label{IW1553n}
\tilde{\rho}_{\mathrm{grav}} \simeq \frac{3}{4} \tau^{2}(\tilde{t}_{\mathrm{J}})  \sim O(1),
\end{equation}
many orders greater than the calculated values $\tilde{\rho}_{\mathrm{rad}} \sim O(10^{-8})$.

This great disparity in the values of $\tilde{\rho}_{\mathrm{grav}}$ and $\tilde{\rho}_{\mathrm{rad}}$ during the slow-roll era 
indicates that the end of reheating and the beginning of radiation dominance,
when $\tilde{\rho}_{\mathrm{grav}} =\tilde{\rho}_{\mathrm{rad}}$, 
lies well into the fast-roll oscillation regime. Using the analytic approximations for the fast-roll regime of the 
$gR^{2}-AB$ model obtained by \cite{MN2012}, this occurs at $t=2.2 \times 10^{16}\tilde{M}$  and a temperature
$T_{\mathrm{RD}} =1.4 \times 10^{7}$ GeV.

It is of interest to note that, whereas most existing calculations of gravitational baryogenesis assume the scale factor dependence 
$a_{\mathrm{J}}(t_{\mathrm{J}}) \sim t_{\mathrm{J}}^{\beta}$, 
the present calculations give, from (\ref{IW822}), (\ref{IW842}) and (\ref{IW1029}), the dependence  
$a_{\mathrm{J}}(t_{\mathrm{J}}) \sim \tau^{1/2}(t_{\mathrm{J}}) \exp\{\frac{3}{4} [d_{\mathrm{init}}^{2}-\tau^{2}(t_{\mathrm{J}})]\}$
for the Starobinsky model, and 
$a_{\mathrm{J}}(t_{\mathrm{J}}) \sim \hat{\tau}(t_{\mathrm{J}})^{(1-\alpha)/(\alpha +2)} 
\exp \{\frac{3}{(\alpha +2)^{2}}[\hat{d}_{\mathrm{init}}^{2} - \hat{\tau}^{\alpha+2}(t_{\mathrm{J}})]\}$ for the power-law  model,
where $\tau \leq d_{\mathrm{init}}$ and $\hat{\tau} \leq \hat{d}_{\mathrm{init}}$, and decrease with $t_{\mathrm{J}}$.

The calculated values $\eta $ for the Starobinsky model vary from $(1.05 - 1.46) \times 10^{-11}$, and, for the power-law model, 
from $(1.06 - 1.53) \times 10^{-11}$.
However, it should be noted that the results for the power-law model
depend upon the values chosen for $c_{1}$ and $c_{2}$, and a future fit of the Odintsov and Oikonomou model to data could yield
enhanced values.

Although the values of $\eta$ for both models are quite close to the observed value $\eta_{\mathrm{obs}} = 8.65 \times 10^{-11}$, 
the calculations cannot be taken as establishing the feasibility of gravitational baryogenesis in these models as they are based 
upon several assumptions that qualify these results. They should only be taken as a \emph{proof of concept}.

The asymmetries are generated during the slow roll inflation era and, for values generated $N$ e-folds before the end of inflation, 
must be diluted by the factor $\Delta \lesssim 10\, e^{-3N}$, making agreement with observation difficult to achieve for these models 
unless baryogenesis occurs at the end of the inflation era.

More fundamental criticisms of the present calculations, raised by the anonymous referee, are the assumption of rapid thermalization
of the gravitationally produced plasma and that the $B$-violating interactions are faster than the expansion rate of the Universe.
Rapid thermalization requires~\cite{KM2011} $\alpha^{2} T > H_{\mathrm{J}}$ where $\alpha =g^{2}/(4 \pi) \sim 7 \times 10^{-3}$ 
for $g^{2} \simeq 0.3$. The slow-roll calculations reported in this study give 
$H_{\mathrm{J}} \sim 0.7 \tilde{M} \sim 2 \times 10^{13}$ GeV and $T \sim 9 \times 10^{12}$ GeV at the end of the slow-roll era, thus strongly violating this assumption.
If it is assumed that the $B$-violating interactions are generated by an operator $\mathcal{O}_{\mathrm{B}}$ of mass dimension 
$D=4+n$~\cite{DKKMS2004}, the rate of such interactions is given by $\Gamma_{\mathrm{B}} =T^{2n+1}/M^{2n}_{\mathrm{B}}$, 
where $M_{\mathrm{B}}$ is the mass scale associated with$\mathcal{O}_{\mathrm{B}}$.
Decoupling of $B$-violating processes occurs when $\Gamma_{\mathrm{B}} > H_{\mathrm{J}}$, requiring 
$M_{\mathrm{B}} \lesssim (T^{2n+1}/H_{\mathrm{J}} )^{1/2n}$. For a dimension-six operator, and taking $H_{\mathrm{J}} \sim 10^{13}$ GeV  
and $T \sim 10^{13}$ GeV for the slow roll era, this requires  $M_{\mathrm{B}} \lesssim (T^{5}/H_{\mathrm{J}} )^{1/4} \lesssim 10^{13}$ GeV. However, GUT or Planck-suppressed interactions require larger values of $M_{\mathrm{B}}$, typically $\geq 10^{14}$ GeV~\cite{DKKMS2004}. 

Validation of the two assumptions requires the Hubble rate $H_{\mathrm{J}}$ to be significantly lower, and the temperature $T$ to be 
significantly higher. Both quantities depend upon the variable 
$\tau (t_{\mathrm{J}})=d_{\mathrm{init}}-\tilde{M}(t_{\mathrm{J}}-t_{\mathrm{J,init}})/3$, and therefore upon the initial value 
$\phi_{\mathrm{init}}=(\sqrt{6}/\kappa)\ln d_{\mathrm{init}}$ of the scalaron field. The calculations have assumed 
$\phi_{\mathrm{init}}=5.6/\kappa$, reflecting the high field value required for the Starobinsky model to be consistent with CMB 
observations. The value $\tilde{M} =1.2 \times 10^{-5} M_{\mathrm{Pl}}$ used is determined from the amplitude normalization of the 
CMB~\cite{MN2012}. As significant variation in these quantities is excluded, it is clear that the present calculations cannot be
varied to validate the two assumptions.

A complete treatment of thermalization of the gravitationally produced plasma, baryon-number violating freeze-out, 
and entropy dilution is required to establish whether or not gravitational baryogenesis during slow-roll inflation is a viable 
mechanism for producing the observed baryon asymmetry of the Universe for these inflationary $f(R)$ cosmologies.


\end{document}